\documentclass[12pt,preprint]{aastex}
\usepackage{epstopdf}
\usepackage{graphicx,rotating}
\usepackage{amssymb}
\usepackage{amsmath}
\usepackage{natbib}
\usepackage{lscape}
\newcommand{\beq}{\begin{equation}}
\newcommand{\eeq}{\end{equation}}
\newcommand{\etal}{{\sl et~al.~}}
\newcommand{\kms}{km s$^{-1}$}
\newcommand{\msy}{mas yr$^{-1}$}
\newcommand{\HST}{{\it HST~}}

\newcommand{\HIP}{{\sl Hipparcos}}

\def\K{$Kepler$~}
\def\Kns{$Kepler$}

\begin{document}

\received{}
\revised{}
\accepted{}

\shorttitle{{\it Kepler} Astrometry}
\shortauthors{Benedict}

\bibliographystyle{/Active/my2}
\title{A Technique to Derive Improved Proper Motions for {\it Kepler} Objects of Interest\footnote{Based on observations made with the NASA {\it Kepler} Telescope}}

\author{G. Fritz Benedict\altaffilmark{2}, \\Angelle M. Tanner\altaffilmark{3}, Phillip A. Cargile\altaffilmark{4},  and David R. Ciardi\altaffilmark{5}}

\altaffiltext{2}{McDonald Observatory, University of Texas, Austin, TX 78712}
\altaffiltext{3}{Department of Physics and Astronomy, Mississippi State University, Starkville, MS 39762}
\altaffiltext{4}{Department of Physics and Astronomy, Vanderbilt University, Nashville, TN 37235}
\altaffiltext{5}{NASA Exoplanet Science Institute, Caltech, Pasadena, CA 91125}



\begin{abstract}
We outline an approach  yielding proper motions with higher precision than exists in present catalogs for a sample of stars in the \K field.   To increase proper motion precision we combine first moment centroids of \K pixel data from a single Season with existing catalog positions and proper motions. We use this astrometry to produce improved reduced proper motion diagrams, analogous to a Hertzsprung-Russell  diagram, for stars identified as {\it Kepler} Objects of Interest. The more precise the relative proper motions, the better the discrimination between stellar luminosity classes. With UCAC4 and PPMXL epoch 2000 positions (and proper motions from those catalogs as quasi-bayesian priors) astrometry for a single test Channel (21) and Season (0) spanning two years yields proper motions with an average per-coordinate proper motion error of 1.0 \msy, over a factor of three better than existing catalogs. 
We apply a mapping between a reduced proper motion diagram and an HR diagram, both constructed using \HST parallaxes and proper motions,  to estimate \K Object of Interest $K$-band absolute magnitudes. The techniques discussed apply to any future small-field astrometry as well as the rest of the \K field.
\end{abstract}


\keywords{astrometry ---  stars: distances --- stars: proper motions   --- stars: exoplanet hosts}


%

\section{Introduction}
Astrometric precision, $\epsilon$, {\it in the absence of systematic error} is proportional to N$^{-1/2}$, where N is the number of observations \citep{WvA13}. 
Theoretically, averaging the existing vast quantity of \K data might permit one to approach \HST/FGS astrometric precision, 1 millisecond of arc per observation. While \K was never designed to be an astrometric instrument, and despite significant astrometric systematics and the challenge of fat pixels (3\farcs9757/pixel), one can reach a particular goal; higher precision proper motions for \K Objects of Interest (KOIs) from \K data. Additionally, these techniques may find utility when future astrometric users of, for example,  the Large Synoptic Survey Telescope \citep{Ive08} require the highest possible astrometric precision for targets of interest contained on a single CCD in the focal plane. Finally, proper motion measures from any \K extended mission might benefit from the application of these techniques.

In transit work it is useful to know the luminosity class of a host star when estimating the size of the companion. This requires a distance, ideally provided by a measurement of parallax. With simple centroiding unaware of point spread function (PSF) structure the season to season \K astrometry required for parallaxes presently yields positions with average errors exceeding 100 milliseconds of arc (mas), insufficient for parallaxes (Section~\ref{WTF2}). Yet distance is a desirable piece of information. Reduced proper motion (RPM) diagrams may provide an alternative distance estimate. The concept is simple: proper motion becomes a proxy for distance \citep{Str39,Gou03,Gou04}. Statistically, the nearer any star is to us, the more likely it is to have a larger proper motion. The RPM diagram consists of the proper motion converted to a magnitude-like parameter plotted against color. 
 The RPM diagram is thus analogous to a Hertzsprung-Russell (HR) diagram. While some nearby stars might have low proper motions, typically giant and dwarf stars are separable. The more precise the proper motions, the better the discrimination between stellar luminosity classes.

In the following sections we describe our approach yielding improved placement within an RPM for any \K target of interest.
 We outline in Section~\ref{RPM} the utility of RPM diagrams, including a calibration to absolute magnitude derived from \HST astrometry; discuss in Section~\ref{OBS} \K data acquisition and reduction; present the results of a number of tests providing insight into the many difficulties associated with \K astrometry (Section~\ref{WTF}); describe the modeling and the proper motion results for our selected \K test field,  (Section~\ref{MOD}); compare our improved RPM with that previously derived from existing astrometric catalogs (Section~\ref{mobet}); and discuss the astrophysical ramifications of our estimated absolute magnitudes for the over 60 KOI's in our test field (Section~\ref{LC}). We summarize  our findings in Section~\ref{SUMM}. 

\section{A Calibrated RPM Diagram}  \label{RPM}
In past \HST astrometric investigations (e.g. Benedict et al. 2011, McArthur et al. 2010)\nocite{Ben11}\nocite{McA10} the RPM was used to confirm the spectrophotometric stellar spectral types and luminosity classes of reference stars. Their estimated parallaxes are entered into the modeling as observations with associated errors.
To minimize absorption effects \HST astrometric investigations  use for the magnitude-like parameter H$_K(0) = K(0) + 5log(\mu)$,  and for a color, (J$-$K)$_0$, where K magnitudes and (J$-$K) colors (from 2MASS, Skrutskie \etal 2006)\nocite{Skr06} have been corrected for interstellar extinction. For all of our RPMs we use the vector length proper motion, $\mu=(\mu_{\rm RA}^2 + \mu_{\rm DEC}^2)^{1/2}$. 

Compared to \HIP, \HST  has produced only a small number of  parallaxes and proper motions (Benedict \etal 1999, 2000a, 2000b, 2002, 2006, 2007, 2009, 2011; Harrison \etal 2013; McArthur \etal 2001, 2010, 2011, 2013; Roelofs \etal 2007), but with higher precision.\nocite{Ben00a} \nocite{Ben06} \nocite{Ben07} \nocite{Ben09} \nocite{Ben07} \nocite{McA10} \nocite{McA11} \nocite{Har13} \nocite{Roe07} Parallax and proper motion results for forty-two stars with \HST proper motion and parallax measures are collected in Table~\ref{THT}. Average parallax errors are 0.2 mas. Average proper motion errors are 0.4 \msy.  Compared in Figure~\ref{HR} are an HR diagram and an RPM diagram constructed with \HST parallax and proper motion results for the targets listed in Table~\ref{THT}. Conspicuously absent from the RPM diagram are the RR Lyr results from \cite{Ben11} with their large proper motions due to their Halo Pop II identification. Lines plotted on the HR diagram show predicted loci for 10 Gyr age solar metallicity stars  and 3Gyr age metal-poor stars from Dartmouth Stellar Evolution models \citep{Dot08}. These ages and metallicities encompass the majority of what might be expected from a random sampling of stars in our Galaxy.

Even though these targets are scattered all over the sky and range from PN central stars to Cepheid variables, the similarity between the HR and RPM diagrams is striking. In Figure~\ref{HRRP} we plot the extinction corrected $K$-band absolute magnitude derived from \HST parallaxes against the magnitude-like parameter H$_K(0)$. The resulting scatter, 0.7 mag RMS,  suggests that precise proper motion is a good enough proxy for distance to allow the assignment of luminosity class. 

Note that while this calibration is produced with proper motions sampling much of the celestial sphere, the HR diagram and RPM main sequences are defined almost exclusively by stars belonging to the Hyades and Pleiades clusters. This may become an issue when we attempt to apply the calibration to a small piece of the sky in a different location. Both the calibration sources and a random \K field  have systematic proper motions due to galactic rotation ({\it c.f.}~van Leeuwen 2007, section 6.1.5), \nocite{Lee07a} possibly requiring some correction.

\section{\K Observations and Data Reduction}  \label{OBS}
The primary mission of the \K spacecraft is high-precision photometry with which to discover transiting planets. \K rotates about the boresight once each 90 days to maximize solar panel illumination. Each such pointing is identified by a Season number; 0--3. Each 90 day span is identified by a Quarter number; 1--17. The CCDs in the \K focal plane are identified by Channel number; 1-84. 
Our goal is to produce an astrometric reference frame across a given \K channel containing KOIs of interest, the end product being KOI proper motions with which to populate an RPM. 

\subsection{Star Data} 
The \K telescope trails the Earth in Sun-centered orbit. Details on the photometric performance and focal plane array can be found in \cite{Bor10} and \cite{Cal10a}. The following explorations restrict themselves to the so-called long-cadence data, where each subsection containing a star of interest in the array is read out once every 30 minutes. These subsections of the \K FOV (hereafter, postage stamps) range from 4x5 pixels for fainter stars to larger than 8x8 pixels for brighter stars. \K pixels are a little less than 4 arc sec on a side. 
Targets observed with long-cadence generate approximately 4700 postage stamps per star per Quarter.

We obtain \K data from the Space Telescope Science Institute Multimission Archive (MAST). These data include both pipelined positions ( the *\_llc.fits files, where `*' is a global replacement marker) and postage stamp image data (the *\_lpd-targ.fits files). The \K Archive Manual \citep{Fra12} greatly assisted us with any access issues.

\subsection{Positions from \K image data} \label{NPG}
Positions used in this paper are generated from the \K postage stamp image data, using a simple first-moment centering algorithm. We calculate 
\begin{eqnarray}
MOM\_CENTR\_X = \sum i*z/\sum i\\
MOM\_CENTR\_Y = \sum j*z/\sum j
\end{eqnarray}
where z = flux(i,j) are the flux count values for each \K pixel within the postage stamp. To generate positions from the optimal aperture, any z value not in the optimal aperture is set to zero. To reduce the computational load and to smooth out high frequency positional variations, normal points (NP) are formed by averaging the x and y positions for a specified time span. The tests and results reported herein are based on nine day normal points. 
We also use only data within an optimal aperture for each star defined by the \K team. We provide an example of an optimal aperture for a star with \K identification number KID = 7031732 in Figure~\ref{OPT}. 
 There are positional corrections tabulated in the MAST data products, e.g. POS\_CORR1. These report the size of the differential velocity aberration
(DVA), pointing drift, and thermal effects applicable to the region of sky recorded in the file. These corrections are applied to our derived centroids.
 Final positions used in the test are corrected using the MAST position correction values, e.g.,  XY\_CORR = MOM\_CENTR\_XY - POS\_CORR\_XY. The standard deviation of each normal point along each axis for each star is calculated. The average standard deviation of these NP is typically on order one mas, demonstrating exceptional astrometric stability within each postage stamp. However. this small formal random error is a significant underestimation of the total star-to-star astrometric quality, as we shall see below in Section~\ref{WTF1}. 

 We explored utilizing PSF fitting methods to extract positions. That approach did not
resolve the issue of poor astrometric performance over multiple quarters (see Section~\ref{WTF2} below). 
PSF fitting is computationally intensive and complicated given the \K FOV crowded stellar field and the
significant PSF variations over that field \citep{Bry10}.

\section{\K Astrometric Tests} \label{WTF}
These tests highlight several systematic errors and motivate our simple  strategy for dealing with them.
For all astrometric modeling we employ GaussFit (Jefferys \etal 1988)\nocite{Jef88} to minimize $\chi^2$. 

\subsection{Single Channel, Single Season, Single Quarter}\label{WTF1}

These  tests use 95 stars  located in Channel 21, Season 0, Quarter 10 and identified as red giants in the \K Input Catalog (KIC). This initial filtering by star type potentially minimizes any effects of proper motion over the span of one or even two Quarters.  The normal point generator, run on each star selected from test Channel 21, effectively reduces the number of discrete data sets per star from on order 4500 down to nine.  We assign the nine normal points for each of the 95 stars in the test to the nine `plates'. Each plate now contains 95 stars whose epochs are now separated in time by approximately nine days. With the positions and positional errors generated by the normal point code  (now organized as nine `plates' each containing 95 star positions and associated errors) we determine scale, rotation, and offset ``plate
constants" relative to an arbitrarily adopted constraint epoch (the so-called ``master plate") for
each observation set (the positions generated for each star  within each `plate' at each of the nine normal point epochs). The solved equations
of condition  are:

\beq
\xi = Ax + By + C 
\eeq
\beq
\eta = -Bx + Ay + F 
\eeq
where $\it x$ and $\it y$ are the measured normal point coordinates from the Kepler postage stamps.
 A and B  
are scale and rotation plate constants, C and F are
offsets. For this test spanning only one 90 day Quarter we ignore proper motions. When modeling these positions, in order to approach a near unity $\chi^2$/DOF (DOF=degrees of freedom), the input positional errors, standard deviations from the normal point averaging process, had to be increased by a factor of four.
The final catalog of $(\xi,\eta)$ positions have average $<\sigma_\xi > = 0.31$ millipixel and $<\sigma_\eta > = 0.64$ millipixel (1.23 and 2.54 mas), seemingly quite encouraging if one's goal is precision astrometry with Kepler. 

However, the results of this modeling, shown in Figure~\ref{Q10r}, exhibit large systematic effects, well-correlated with time.  The constraint epoch for this reduction is JD-24400000=15785.2, the middle epoch of the nine plotted. We tentatively blame the typically larger y residuals (y larger than x within each epoch) in Figure~\ref{Q10r} with charge transfer smearing along the CCD column readout direction \citep{Koz08,Qui10}. Our ultimate goal is to tease out stellar positional behavior similarly correlated with time; proper motion. Among the largest and most strikingly systematic residual patterns we find stars 9 (=KID 6363534) and 27 (=KID 6606001). Figure~\ref{FUNS} provides an explanation for the behavior of star 27 (a close companion that perturbed the simple first moment centering algorithm), and presents a puzzle regarding Star 9. It has no bright companions, yet is a poorly behaved component of our astrometric reference frame. We suspect the CCD channel-to-channel cross-talk discussed in \cite{Cal10b}. Four CCDs share common readout electronics. A bright star on one CCD can affect measured charge in another.

To determine if there might be unmodeled - but possibly correctable -  systematic effects at the 10 millipixel level, we plotted the reference frame x and y residuals against a number of  parameters. These included x, y position within the channel; radial distance from the channel center; reference star  magnitude and color; and epoch of observation.  We saw no obvious trends, other than an expected increase in positional uncertainty with reference star magnitude. Models with separate x and y scales (6 parameters; in Equation 2 above, where -B and A are replaced by D and E) or color terms (8 parameters) provided no improvement in $\chi^2$/DOF.

Given that the pipelined positions available in the \_llc.fits files are also first moment centroids calculated from the optimal apertures, we developed the capability to generate these independently as a further test of \K astrometry. The code used to generate the positions whose residuals are plotted in Figure~\ref{Q10r} can also produce first moment centroid data using the entire postage stamp (e.g., all the flux values in the middle panel in Figure~\ref{OPT}). Comparing positions extracted from the entire postage stamp against the optimal subset of the postage stamp for Channel 21, the average absolute value residual is reduced by 30\% when using the optimal apertures. However tests carried out on Channel 41 (Season 0, Quarter 10), near the \K field of view center result in a much less significant improvement, only 12\%. This can be explained by considering the degradation in the  point spread function (PSF) from the center to the edge of the entire \K field of view \citep{Bry10,Ten10}.

\subsection{Other Single Channel Tests}
\subsubsection{Same Stars - Multiple Seasons and Quarters}
To test whether or not the peculiar fan pattern in the residuals against time seen in Figure~\ref{Q10r} is Channel-specific, we carried out similar tests for Channels 41, 42, 43, and 44, the central four CCDs in the \K focal plane. We sampled Quarter 3 through Quarter 14, using the same $\sim$75 stars in each channel. The results of this 4 parameter modeling are given in Figure~\ref{ALLQ}. Every Quarter exhibits time-depends residual behavior.  The patterns often repeat within the same season. For example, in Season 1 the x residuals for star 5 (=KID 8949862)  start out large and positive and move down to large negative over $\sim90$ days. Yet star 5 is relatively well-behaved for any other Season. A comparison of Figure~\ref{ALLQ} with figure 5  in \cite{Bar11}\footnote{http://archive.stsci.edu/kepler/release\_notes/release\_notes12/\\DataRelease\_12\_2011113017.pdf} is convincing evidence that astrometry quality and primary mirror temperature changes are correlated. Stable temperatures at any level yield better astrometry (smaller residuals). 

\subsubsection{An External Check of Single Channel, Single Season Data}
We carried out  a four parameter modeling of 127 stars randomly chosen (a mix of dwarfs and giants according to the KIC) in Channel 26 from Season 3, Quarter 5 and found residual patterns similar to that for Season 3, Quarter 5, Channel 44 shown in Figure~\ref{ALLQ}. We then extracted a subset of ten stars, five with relatively well-behaved x residuals (stars 4--62), five with x residuals not constant with time (stars 67--106 in Figure~\ref{GBres}). We list these in Table~\ref{Textern}.

To explore the hypothesis that astrophysical effects (i.e., companions undetectable at the resolution of the \K detectors) cause the observed residual behavior, these ten stars were observed with the Keck NIRC2-AO system (Wizinowich \etal 2004; Johansson \etal 2008)\nocite{Wiz04,Joha08} on the nights of 19-21 August 2013 UT with the NIRC2 instrument on Keck II.  The targets themselves were used as natural guide stars and observations were made in the K' filter or the Br-$\gamma$ filter, if the star was too bright for the broader K' filter. The native seeing on the three nights (before AO) was approximately 0\farcs6 at 2$\mu$m.  The NIRC2 instrument was in the narrow field mode with a pixel scale of approximately 0\farcs009942 pixel$^{-1}$ and a field of view of approximately 10\farcs1 on a side.  Each dataset was collected with a 3-point dither pattern, avoiding the lower left quadrant of the NIRC2 array, with 5 images per dither position, each shifted 1" from the previous. Each frame was dark subtracted and flat fielded. The sky frames were constructed for each target from the target frames themselves by median filtering and coadding the 15 dithered frames. Individual exposure times varied depending on the brightness of the target but typically were 10 - 30 seconds per frame.  Data reduction was performed with a custom set of IDL routines.


To estimate companion detection limits as a function of distance from the selected targets we utilized PSF planting and cross correlation (Tanner et al. 2010)\nocite{Tan10}. The science target was extracted from the image, sky subtracted, and normalized. Then it was added at random positions around the image such that an equal number of plant positions occur in each radius bin of 0\farcs1 around the science target. The flux within each planted PSF is scaled by a random value which ranges from $10^{-3 }$ to $10^3$ times the original number of counts in the star. The counts were determined through aperture photometry with a radius of 0\farcs2 and a sky annulus of 0\farcs2-0\farcs3. Once added to the image, the threshold for detection was established by cross-correlating the planted star with the normalized PSF. A scaled and random PSF plant was considered detected if the cross correlation value was above 0.5 - a value determined by a "by eye" assessment. The location and flux of those PSFs which were detected were recorded over 5000 PSF plants. In each radius bin the PSF with the smallest flux was used in the resulting plot of detected minimum magnitude difference ($\Delta K_s$) versus distance from the science target.  

We found no companion candidates in these images within a radius of 1\farcs2. Figure~\ref{ConGB} contains the contrast curves for star 4 (constant residuals) and star 67 (time-varying residuals), along with a fit to normalized average contrast curves for stars 4--62 ($\langle \rm Good \rangle$) and 67--106 ($\langle \rm Bad \rangle$). To fit the average contrast curves we employed an exponential function ($y = K0+K1*exp(-(x-x0)/K2)$) with offset, x0. The similarity of the average contrast curves removes small angular separation, fainter companions as the cause of the Figure~\ref{GBres} behavior. Finally, the average Season 3 crowding, contamination, and flux fraction parameters (see Fraquelli and Thompson 2012 for parameter details) of the two groups differed little.

\subsubsection{Lessons Learned}
With as rich a set of data as produced by \K our approach is to exercise extensive editing to establish the best astrometric reference frame; a reference frame with $\chi^2$/DOF$\sim$1, and Gaussian distribution of residuals. If we model only epochs 3--7 in Figure~\ref{Q10r}, as shown in Figure~\ref{Q10ed}, we generate a final catalog of $(\xi,\eta)$ positions with average $<\sigma_\xi > = 0.22$ millipixel and $<\sigma_\eta > = 0.46$ millipixel (0.87 and 1.83 mas). The residuals are Gaussian (Figure~\ref{Q10H})  and naturally larger than the average catalog errors because of the effective averaging to produce a catalog. Again, the significantly larger residuals along the y axis are likely due to CCD read-out issues \citep{Koz08,Qui10}.

\subsection{Two Channels, Two Contiguous Seasons}\label{WTF2}

Again restricting our test to include only stars identified as red giants to minimize proper motion effects, we now run a plate overlap model for the same set of stars appearing on two different Channels (21, 37) for Quarters 10 and 11 respectively. Given that the average absolute value UCAC4 proper motion for this suite of test stars is 7.5 \msy (less than 2 millipixel yr$^{-1}$), the roughly 180 day span of these data should exhibit very little scatter due to unmodeled motions. A four parameter model (with the constraint plate chosen to be from Channel 21) yielded a final catalog with $<\sigma_\xi > = 2.60$ millipixel and $<\sigma_\eta > = 6.6$ millipixel (10.33 and 26.23 mas), significantly poorer astrometric performance than for a single Channel and Quarter (Section~\ref{WTF1}). A six parameter model with separate scales along x and y yielded only a 0.8\% reduction in the large value of reduced $\chi^2$/DOF.

The \K telescope has a Schmidt-Cassegrain design. An effective astrometric model for such a telescope, used successfully in the past on Palomar Schmidt photographic plates, is introduced in \cite{Abb75} and used in e.g. \cite{Ben78}. That model,
\beq
\begin{split}
\xi = Ax' + By' + Cx'y' + Dx'^2 + Ey'^2 \\
+Fx'(x'^2+y'^2) + G    
\end{split}
\eeq

\beq
\begin{split}
\eta = A'x' + B'y' + C'x'y'  + D'x'^2 + E'y'^2\\
 +F'y'(x'^2+y'^2) + G' 
\end{split}
\eeq 
when applied to the Channels 21 and 37 data provided a 9.1\% reduction in reduced $\chi^2$/DOF, but a final catalog with errors almost exactly as found for the four and six parameter models. The run of residual with time is shown in Figure~\ref{C21C37}. The residuals from this modeling are on average eight times larger than for Channel 21 alone. That the residuals remain large even with a Schmidt model demonstrates that the astrometric effects are not due to the Schmidt nature of the \K telescope. The residuals as a function of position within the Channel 37 CCD show large variations on extremely small spatial scales (Figure~\ref{37res}). We have yet to identify the source of these high-frequency spatial defects, but cannot yet rule out the individual field flatteners atop each module containing four Channels \citep{Ten10}. This inter-Channel behavior effectively prohibits the measurement of precise parallaxes using only \K data.

\section{Astrometry of a \K Test Field} \label{MOD}
Our ultimate goal is to produce an RPM diagram including KOIs, permitting an estimate of their luminosity class. This may be feasible by restricting astrometry to a single Channel and Season. Essentially we may be able to ignore the deficiencies demonstrated in Figures~\ref{C21C37} and~\ref{37res} because each star in any given Season will be observed by the same pixels, and the starlight is passing through the exact same region of the field flattener. The seventeen available Quarters provide 3--4 same-Season observation sets for any \K Channel. Examination of Figure~\ref{ALLQ} supports our selection of Season 0 as one of the more stable. Tests similar to those carried out in Section~\ref{WTF1} yielded very poor results for Quarter 2, hence it is unused here.

\subsection{Populating an RPM Diagram} \label{DaList}
To fully populate the H$_K$, (J-K)$_0$ plane of an RPM diagram we extract  Channel 21 long-cadence image data (*lpd-targ*) for stars with 14 $>$KEPMAG $>$ 11.6; Quarters 6, 10, and 14, all Season 0:
\begin{enumerate}
   \item      $\sim 100$ stars classified as red giants,
  \item       $\sim 100$  stars with T$_{\rm eff }>$ 6500K ,
  \item      $\sim 100$  stars with 6200 $>$ T$_{\rm eff } >$ 5100K ,
  \item     $  \sim 30$ stars with T$_{\rm \rm eff} <$5000K and Total\_PM $\geq 0\farcs24 yr^{-1}$ with any KEPMAG value, and
 \item        all KOIs found in Channel 21; e.g., {\it Planetary\_candidate} and {\it Exoplanet\_host\_star} condition flag objects. These, too, have unrestricted KEPMAG.
\end{enumerate}
We extract positions and generate 9-day average normal points using only the \K team defined optimal apertures for each star. When including these data in our modeling with the ground-based catalogs, the \K x,y values are re-origined to the center of  Channel 21.

\subsection{Reference Star Priors}
To place our relative astrometry onto a Right Ascension, Declination system we extract J2000 positions and proper motions from the UCAC4 \citep{Zac13}  and PPMXL \citep{Roe10} catalogs. The catalog positions scale the \K astrometry and provide approximately a 12 year baseline for proper motion determination. Additionally, the catalog proper motions with associated errors are entered into the modeling as quasi-Bayesian priors. These values are not entered as hardwired quantities known to infinite precision. The $\chi^2$ minimization is allowed to adjust the parameter values suggested by these data values within limits defined by the data input errors. 

The input positional errors average 19 mas for the UCAC4 and 63 mas for the PPMXL. The average per axis proper motion errors are 2.6 \msy~for UCAC4 and 3.9 \msy~for PPMXL. A comparison of the two catalogs yields an average per star absolute value proper motion disagreement of 5.1 \msy, indicating room for improvement. The RA and Dec positions from the two catalogs are used to calculate $\xi, \eta$ standard coordinates transformed from radians to seconds of arc \citep{vdK67}, using the center of Channel 21 as the tangent point.

\subsection{The Proper Motion Astrometric Model}\label{bmod}
With the central five epochs of positions from Quarter 6, Quarter 10, and Quarter 14  from \K Channel 21 (the editing of each Quarter illustrated by comparing Figure~\ref{Q10r} to Figure~\ref{Q10ed}), spanning 2.14 years, standard coordinates from PPMXL, and UCAC4, and proper motion priors from the latter two catalogs we determine  ``plate
constants" relative to the UCAC4 catalog (it having smaller formal positional errors). The  constraint epoch is thus 2000.0.  Our reference frame after pruning out astrometrically misbehaving  objects contains 226 stars. For this model we include only those stars with a restricted magnitude range, 14 $ > $ KEPMAG $>$ 11.6, samples 1-3 discussed in Section~\ref{DaList} above. The average magnitude for this magnitude-selected reference frame is $\langle KEPMAG\rangle= 13.3$.

Again, we employ GaussFit (Jefferys \etal 1988)\nocite{Jef88} to minimize $\chi^2$. The solved equations
of condition for the Channel 21 field are now:
\beq
\begin{split}
\xi = Ax' + By' + Cx'y' + Dx'^2 + Ey'^2 \\
+Fx'(x'^2+y'^2) + G  - \mu_x'  \Delta t  
\end{split}
\eeq

\beq
\begin{split}
\eta = A'x' + B'y' + C'x'y'  + D'x'^2 + E'y'^2\\
 +F'y'(x'^2+y'^2) + G'  - \mu_y'\Delta  t  
\end{split}
\eeq

\noindent where $\it x'=x -500$ and $\it y'=y-500$  are the re-origined measured coordinates from \K and the standard coordinates from UCAC4 and PPMXL;
$\mu_x$ and $\mu_y$ are proper motions; and $\Delta$t is the epoch difference from the mean epoch.

From the resulting astrometric parameters we form a plate scale 
\beq
S=(BA'-AB')^{1/2}
\eeq
and find for the 15 epochs (five for each of the three Quarters) of Kepler observations $\langle$S$\rangle$= 3\farcs97664$\pm$0\farcs000009 pixel$^{-1}$, close to the nominal \K plate scale \citep{Cal09} and an indication that the Kepler telescope plate scale as sampled in Channel 21 was quite constant. The scale factor of the PPMXL catalog relative to the UCAC4 catalog was 1.000012.

\subsection{Assessing the Reference Frame}
Using the UCAC4 catalog as the constraint plate, to achieve a  $\chi^2$/DOF $\sim$1 the  \K normal point data errors (normal point standard deviations) had to be increased by a factor of sixteen.  Histograms of the \K normal point residuals were characterized by $\sigma_x=3.6$ mas, $\sigma_y=6.4$ mas. The average absolute value residual for  \K was 4.8 mas; for UCAC4, 24.3 mas; for PPMXL, 61.6 mas. The resulting 226 star reference frame `catalog' in $\xi$ and $\eta$ standard coordinates  was determined
with	average positional errors $\langle\sigma_{\xi,\eta}\rangle= 8.6$ mas, a 55\% improvement in relative position over the UCAC4 catalog. The average proper motion error for the stars comprising the reference frame is 0.8 \msy. 

Again, to determine if there might be unmodeled - but possibly correctable -  systematic effects, we plotted reference frame x and y residuals against a number of parameters. These included x, y position within the channel (Figure~\ref{Ch21res}); radial distance from the channel center; reference star magnitude and color; and epoch of observation.  We saw no obvious trends, other than an expected increase in positional uncertainty with reference star magnitude. Plots of x and y residual versus pixel phase also indicated no trends. We calculate pixel phase; $\phi_x=x-int(x+0.5)$, where $int$ returns the integer part of the (for example) x coordinate.

\subsection{Applying the Reference Frame}\label{finpm}
To insure that the typically fainter (hence less valuable contributors to the astrometric reference frame) stars do not affect our astrometric modeling of the Channel 21 CCD, the identical model in Section~\ref{bmod} is re-run adding normal point positions for the K and M stars (sample 4) and KOI's (sample 5) from Section~\ref{DaList}, holding the equations 7--8 coefficients A--G' to the values determined in Section~\ref{bmod}. Note that we do solve for positions and proper motions. This yields final catalog positions and proper motions for 301 stars representing all the categories listed in Section~\ref{DaList}. The inclusion of fainter stars results in a `catalog' with $\xi$ and $\eta$ standard coordinates average relative positional errors $\langle\sigma_{\xi,\eta}\rangle= 11.1$ mas, and an average proper motion error for all stars of 1.0 \msy. The decrease in proper motion precision relative to that found for the reference frame-only stars is driven by the inclusion of the typically fainter K, M, and KOI stars with average $\langle KEPMAG\rangle= 15.1$. As shown in Section~\ref{AstD} below, the centroids of fainter stars have lower signal to noise, and, if included, would degrade our astrometric reference frame.

We present final KOI proper motions and errors  in Table~\ref{KOIpm}. These are in a sense absolute proper motions, because of the use of prior information. To reiterate, we treated the UCAC4 and PPMXL proper motion priors  as observations with corresponding errors. The Table~\ref{KOIpm} KOI proper motion parameters (and those for the entire set of reference stars modeled above) were adjusted by various amounts depending on the data input errors to arrive at a final result that minimized $\chi^2$.

\subsection{Reference Star Photometric Stability}
Our normal point generation process (Section~\ref{NPG}) also produces an average magnitude. In the case of nine-day normal points, all the measured flux values in each optimum aperture are averaged over the nine day interval and converted to a  magnitude with an arbitrary zero-point through $m_f = 25.768 - 2.5*log_{10}(\langle flux\rangle)$. Because we restricted this test to a single channel, no background correction is applied. The standard deviation for the 15 average $m_f$ magnitudes is plotted against reference star ID number in Figure~\ref{PhotV}. Referencing Section~\ref{DaList}, stars 1--99 are classified as red giants in the KIC (sample 1) ; stars 100--199 are hotter stars (sample 2); stars 201--299 are intermediate temperature stars (sample 3); stars 300--350 are selected to be more likely K and M dwarfs (sample 4); and stars 400 -- 452 are the KOI's found in Channel 21 (sample 5). Note that all ID numbers are not present in the plot due to the editing process (Section~\ref{bmod}) producing the final astrometric reference frame. 

Highest maximum variability with a nine-day cadence is found in our sample of suspected giants, not an unexpected result \citep{Bas13}. The KOIs seem to have photometric variability characteristics most similar to the K,M group. We note the trends to smaller variation with increasing number within each group (as defined in Section~\ref{DaList}). This may be a function of photometric noise characteristics having positional dependence within Channel 21. The selection process populating each group and allocating a running number within each group, always assigned the lowest numbers nearer (x,y)=(0,1000), the highest nearer (x,y)=(1000,0).

\subsection{Astrometry as a Diagnostic}\label{AstD}
Figure~\ref{xA} contains an average absolute value \K x residual for each reference star and KOI (numbering from Table~\ref{KOIpm}) as a function of $m_f$. The residuals are calculated from the Section~\ref{bmod} modeling results. We choose the x residual as a potential diagnostic given that the y residuals in general are systematically larger (c.f. Figures~\ref{Q10H} and~\ref{Ch21res}). The trend line is a quadratic fit to the x residuals for the reference stars only (samples 1--3 in Section~\ref{DaList}). The planet-hosting KOI over-plotted with large font show no extreme astrometric behavior, all lying within $\pm$3-$\sigma$ of the relation. However several KOI hosting unconfirmed planetary companions exhibit astrometric peculiarities. Both KOI 426 and 452 were inspected in 2MASS and Palomar Sky Survey images and showed no nearby stellar companions or image structure indicative of close stellar companions. In addition KOI 452 is the most photometrically variable (Figure~\ref{PhotV}) host candidate star.  Unfortunately, given the random eruption of astrometric peculiarity (c.f. Figure~\ref{FUNS}), astrometry alone cannot serve as a reliable indicator of astrophysically interesting behavior.

\subsection{RPM Diagrams; Pre- and Post-\K}\label{mobet}
We now have the proper motions  required to generate  H$_K(0)$ values for an RPM (Section~\ref{RPM}). $K$-band magnitudes, J$-$K colors and interstellar extinction values, A$_V$, E${(B-V)}$ were extracted from the online Kepler target data base at the MAST.  We  assumed \citep{Sch98}extinction-corrected by K(0) = K-A$_K$, $(J-K)_0$ = $(J-K)$-E${(J-K)}$, with A$_K$ = A$_V$/9 and  E${(J-K)} $= 0.53*E{$(B-V)$}.  The left-hand RPM in Figure~\ref{RPMs} shows H$_K(0)$ and  $(J-K)_0$ for all stars except the KOIs and exhibits a distribution of points that appears to have a main sequence and an ascending sub-giant branch. The average H$_K(0)$ error is 0.43 mag, but is dependent on the value of $\mu_{\rm vec}$ with an increased error towards bright values of H$_K(0)$. 

The scatter in Figure~\ref{RPMs}, left, can be due to several causes. These include proper motion accuracy, random motions of stars, and systematic motions of stars. The H$_K(0)$ average error bar in the figure indicates $\pm0.4$ magnitude of scatter due to measurement error.  The amount due to random stellar motions is unknown. Any particular star could have a large radial component to its random motion and be erroneously placed amongst the giant stars with typically lower than average proper motions. Those two effects increase the random scatter in an RPM. That systematic motions can corrupt an RPM is illustrated in \cite{Ben11}, figure 3. There the RR Lyr variables all lie below and blueward of the broad main sequence. These Pop II giant stars have anomalously large proper motions, causing their erroneous placement in the RPM.

\begin{figure}
\includegraphics[width=\textwidth]{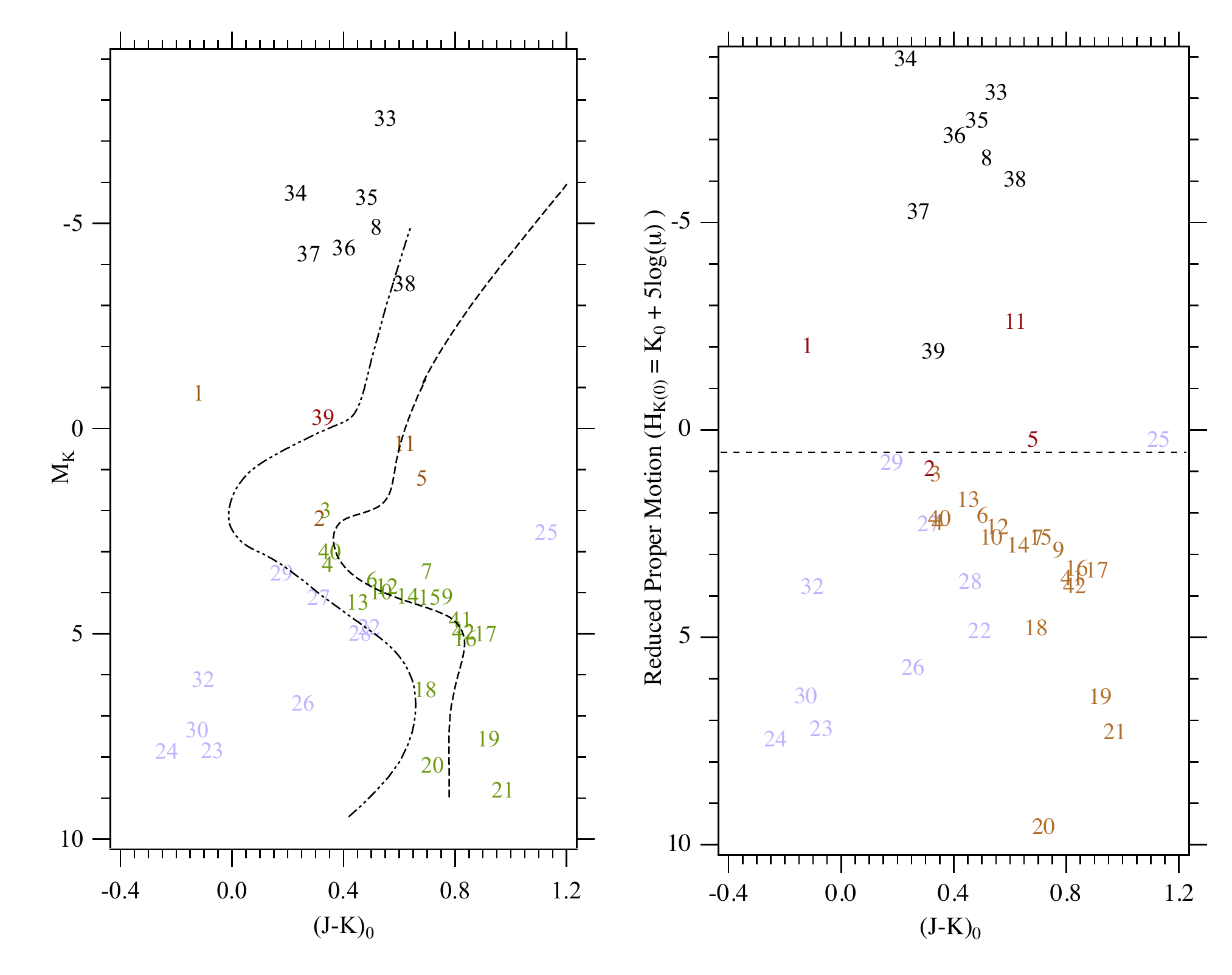}
\caption{Left: Hertzsprung-Russell (HR) diagram absolute magnitude M$_K$(0) vs (J-K)$_0$, both corrected for interstellar extinction. Plotted lines show predicted loci for 10Gyr age solar metallicity stars (- - -) and 3Gyr age metal-poor ([Fe/H]=-2.5) stars (-$\cdot\cdot$-) from Dartmouth Stellar Evolution models \citep{Dot08}.  Right: RPM diagram, same targets plotted. Horizontal line separates main sequence and sub-giants, giants and super-giants. Star numbers are from Table~\ref{THT}. Color coding denotes main sequence (red), white dwarfs (blue), super-giants (black).
 }
\label{HR}
\end{figure}

\begin{figure}
\includegraphics[width=5in]{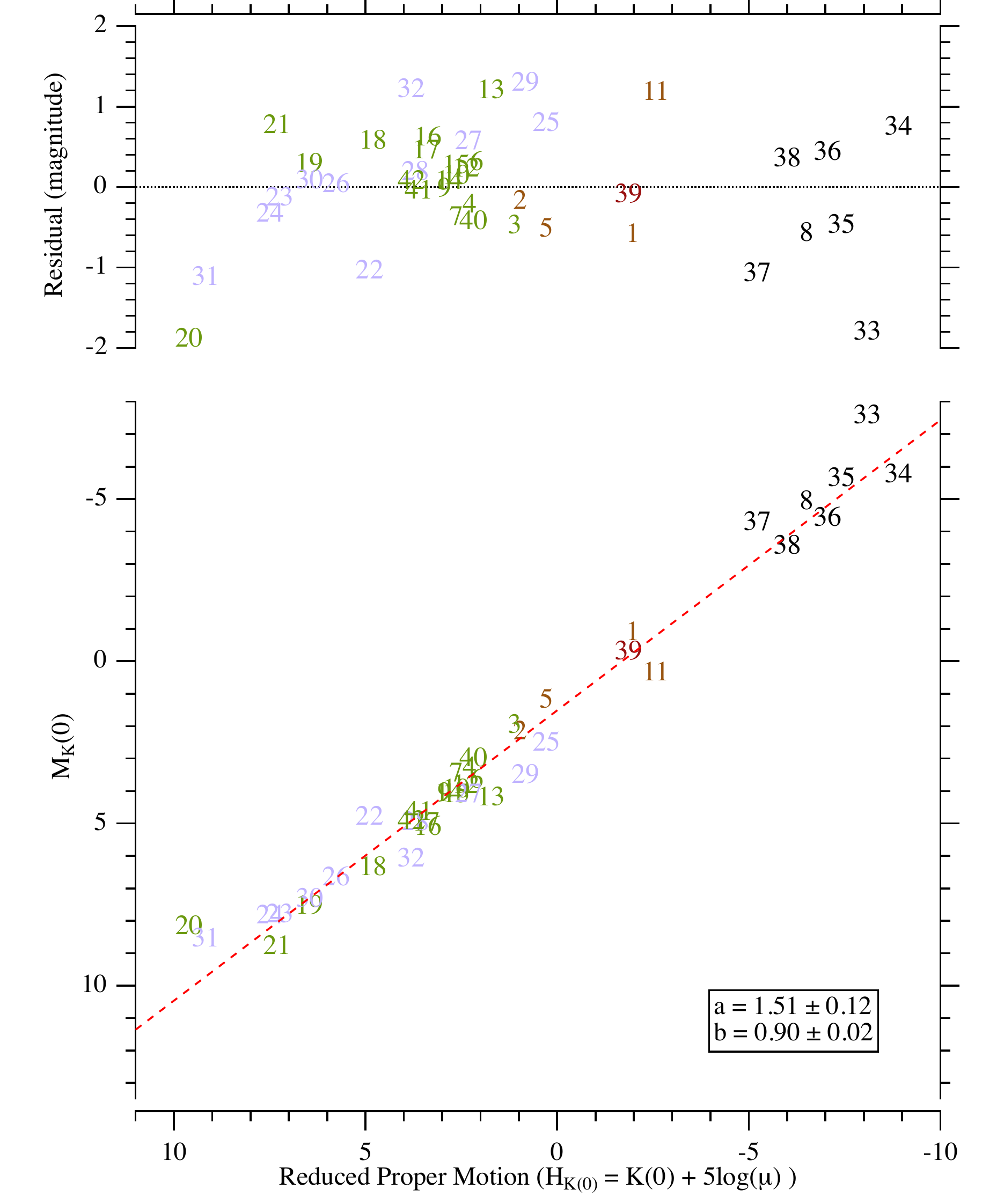}
\caption{A linear mapping between M$_K(0)$ and H$_K(0)$ using \HST parallaxes and proper motions for targets scattered over the entire sky. RMS residual is 0.7 mag. The linear fit (M$_K(0)$ = a + b H$_K(0)$) coefficient errors are 1-$\sigma$. Stellar classifications range from white dwarfs to Cepheids, as listed in Table~\ref{THT}.
}
\label{HRRP}
\end{figure}

\begin{figure}
\includegraphics[width=7in]{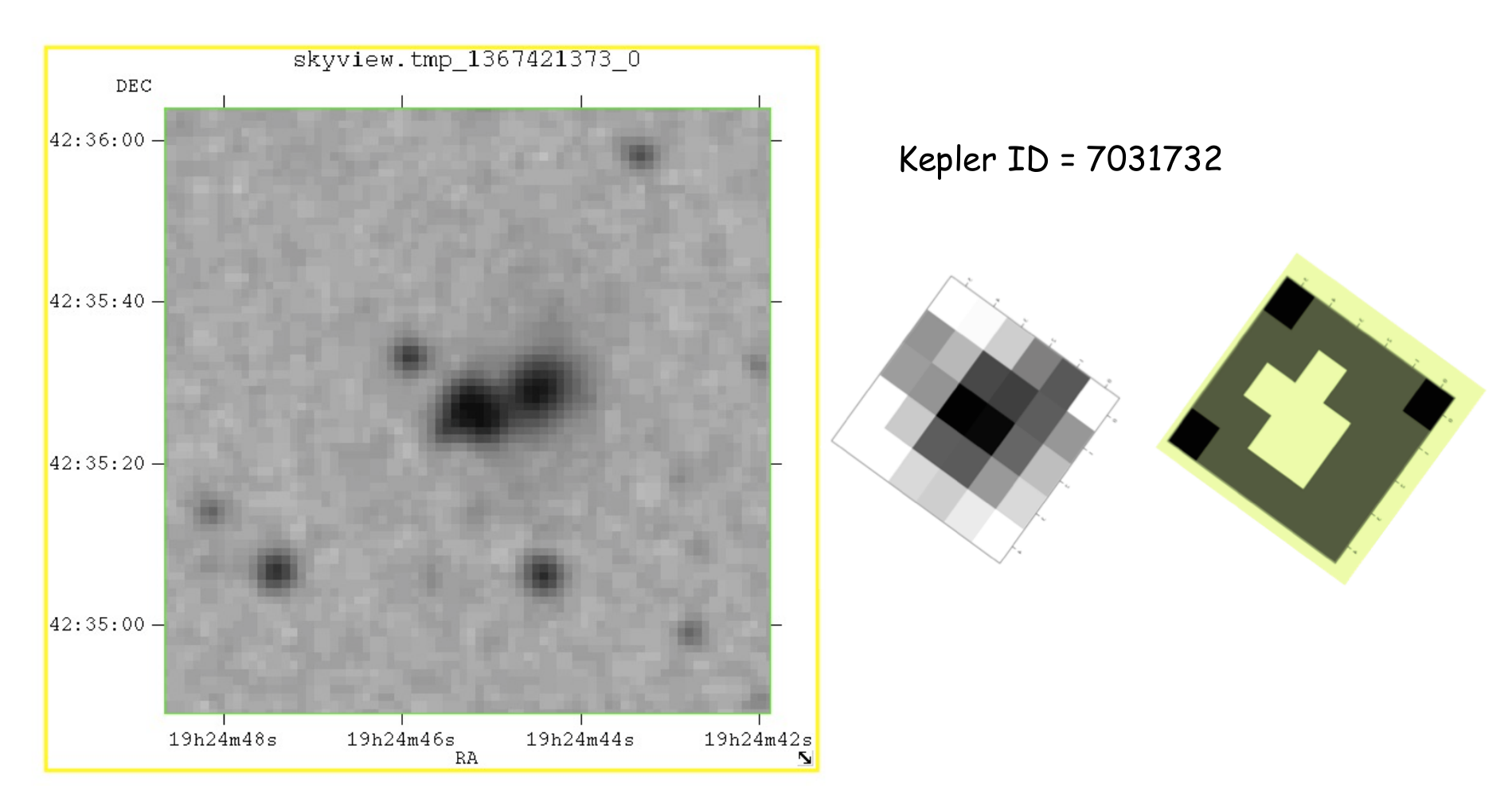}
\caption{Left: KID 7031732 in a crowded field. Image from Digital Sky Survey via {\it Aladin}. Middle: Postage stamp for KID 7031732. Right: Optimum aperture for KID 7031732.}
\label{OPT}
\end{figure}

\begin{figure}
\includegraphics[width=6in]{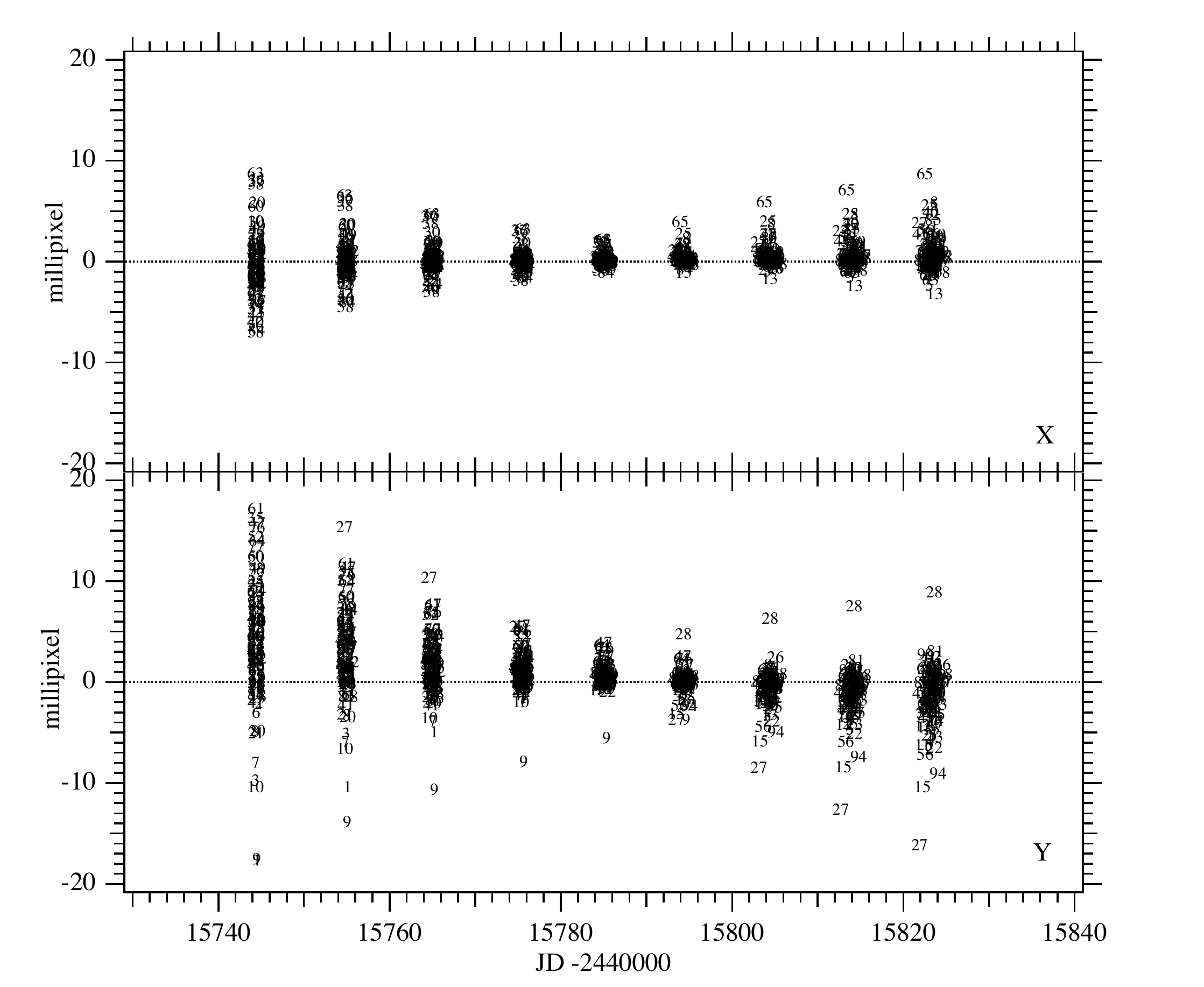}
\caption{x and y residuals as a function of time for the Q10-only four parameter modeling from Section~\ref{WTF1}. The residual clumps from left to right are 'plates' 1--9, the epochs of the averaged normal points. Stars are labeled with a running number from 1 to 97. The residuals exhibit significant time-dependency. Regarding two of the stars with more extreme behavior, neither star 9  (=KID 6363534) nor star 27 (=KID 6606001) is a high-proper motion object.}
\label{Q10r}
\end{figure}

\begin{figure}
\includegraphics[width=6in]{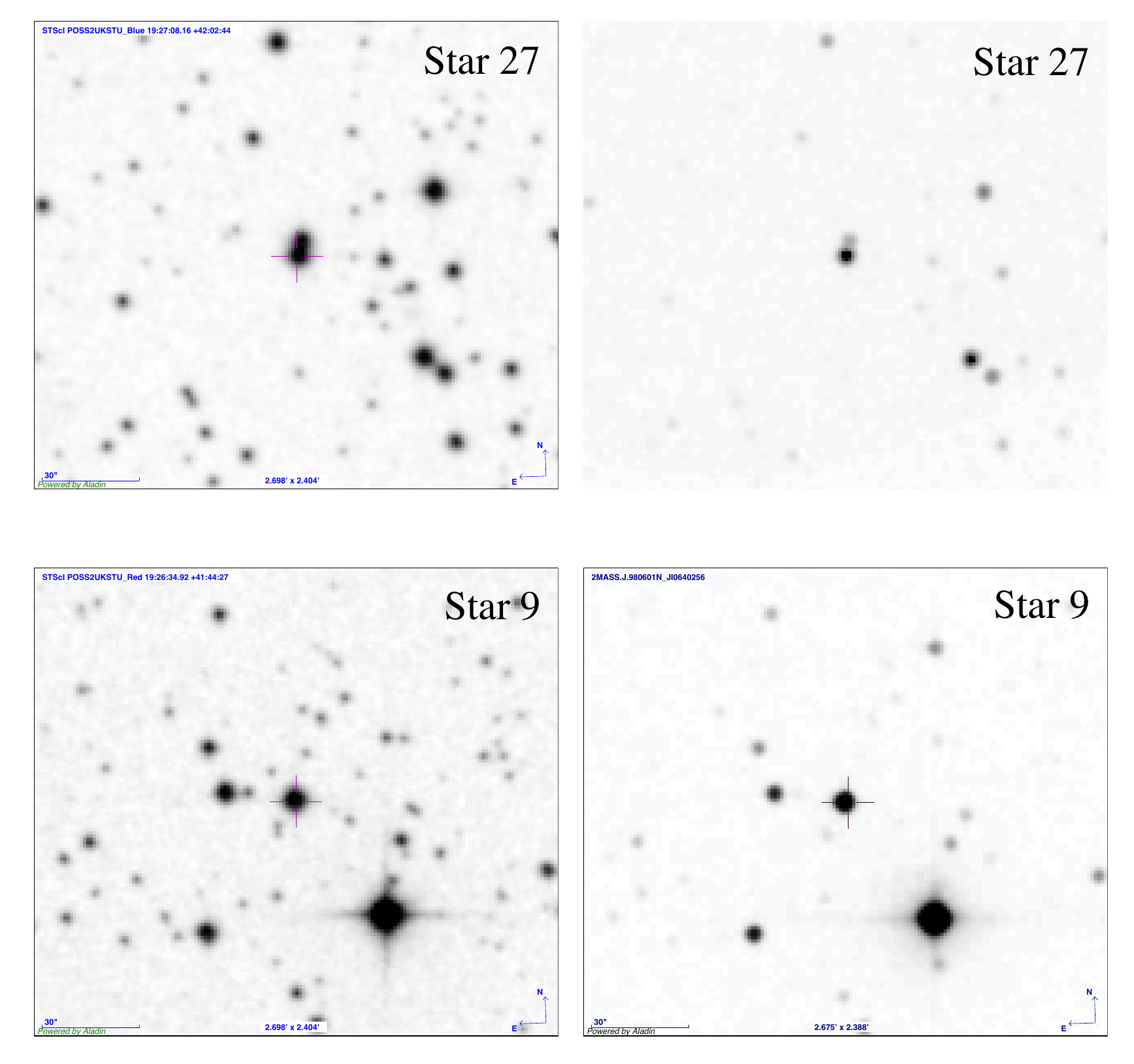}
\caption{Top: Star 27 (POSS-J on left, 2MASS on right) obviously with a close companion that confused the first moment centering. Bottom: Star 9, similarly illustrated. No companion to Star 9 is detected. The positional shift is assumed instrumental. }
\label{FUNS}
\end{figure}

\begin{figure}
\includegraphics[width=7.5in]{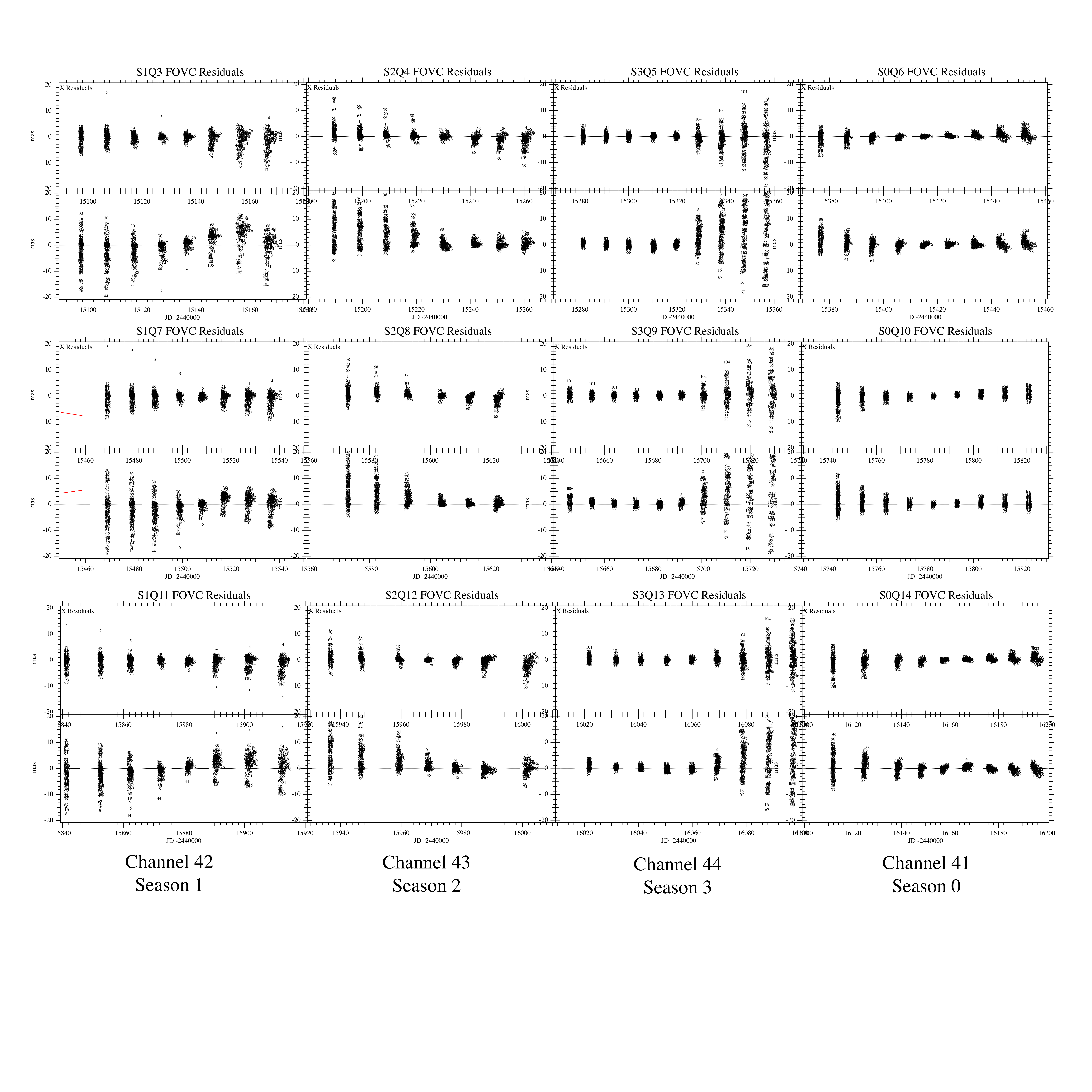}
\caption{As in Figure~\ref{Q10r}, x and y residuals as a function of time for four parameter modeling of a sample of KIC-identified giants in Channels 41--44, but for 11 Quarters. Top row, Quarters 3--6; middle row, Quarters 7--10; bottom row, Quarters 11--14.  Top half of each box contains x residuals; y  below. The y-axis range within each coordinate half-box is $\pm$20 millipixels with an x-axis range between 80 and 90 days. The residuals exhibit  time-dependency within each Quarter that correlates with focal plane temperature changes. Season 0 appears to have a larger fraction of stable astrometry.}
\label{ALLQ}
\end{figure}

\begin{figure}
\includegraphics[width=6in]{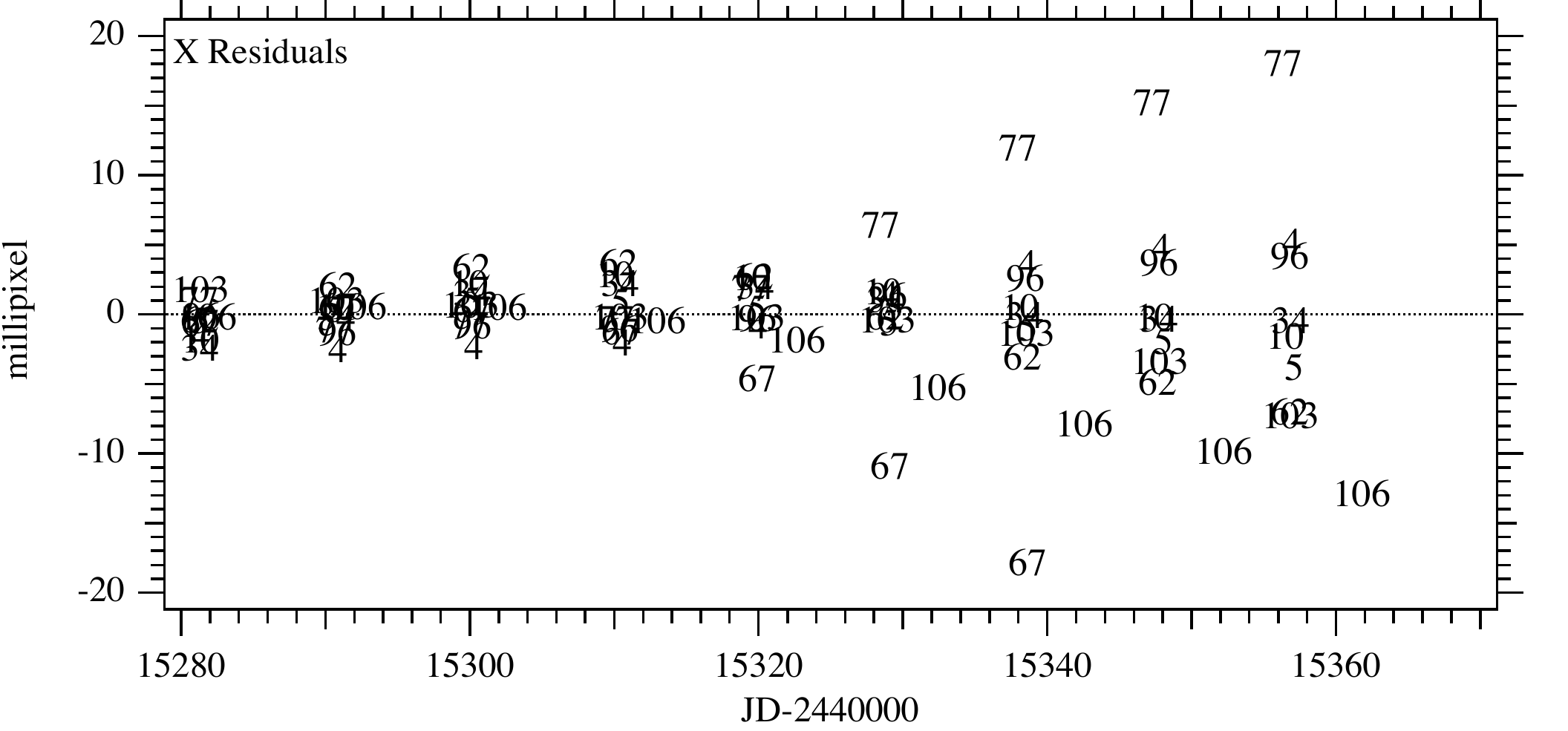}
\caption{Selected x residuals as a function of time for a four parameter modeling (Equations 3--4) of 127 stars in Channel 26 from Season 3, Quarter 5. The star numbers are identified with \K IDs in Table~\ref{Textern}. Note that stars [4, ..., 62] are more astrometrically stable than stars [67, ..., 106].}
\label{GBres}
\end{figure}

\begin{figure}
\includegraphics[width=6in]{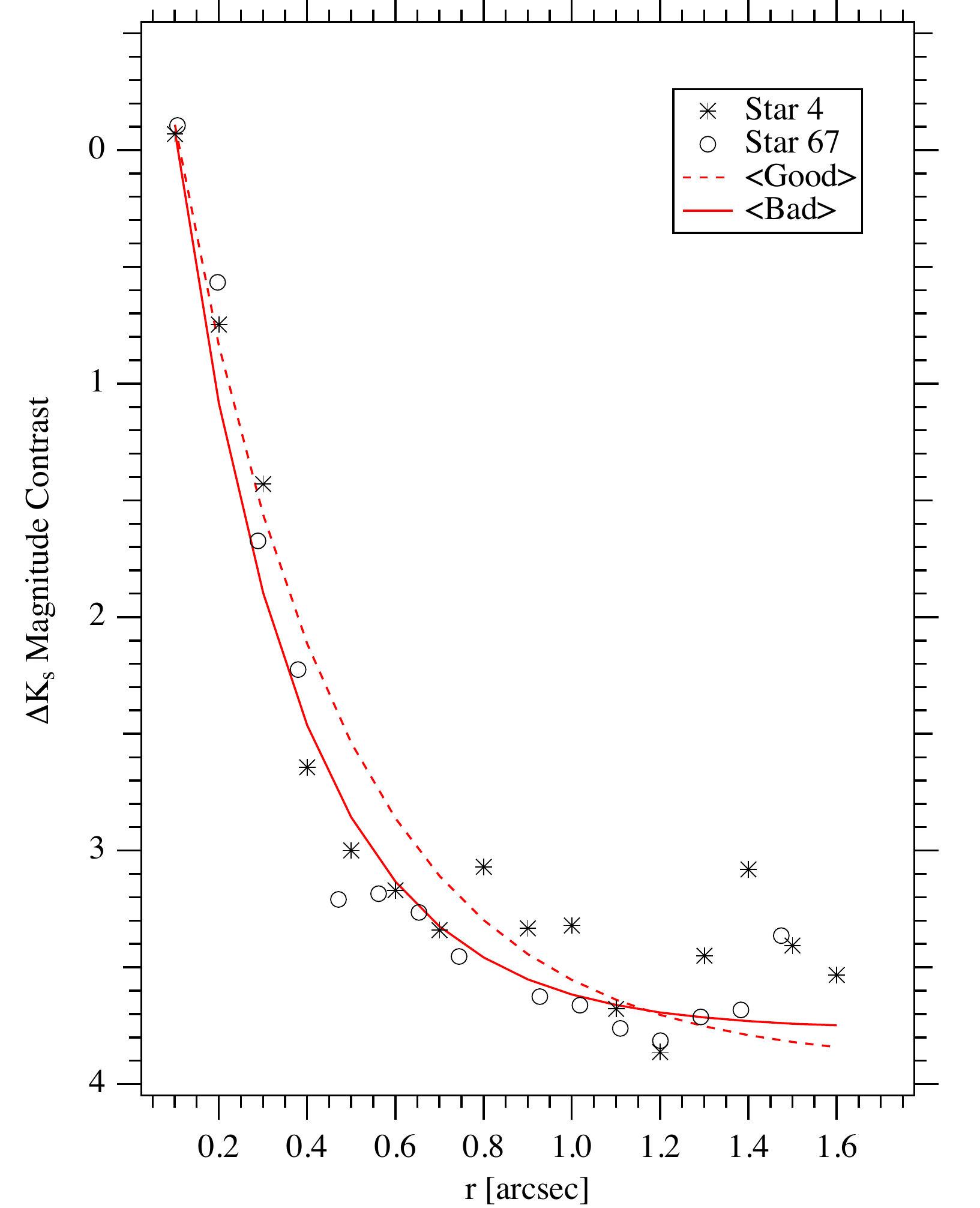}
\caption{Normalized $K$-band contrast curves for stars 4 and 67, along with average contrast curves for stars 4 through 62 (Good) and 67 through 106 (Bad). Note that while star 4 is more astrometrically stable than star 67, they have virtually identical contrast curves.}
\label{ConGB}
\end{figure}

\begin{figure}
\includegraphics[width=6in]{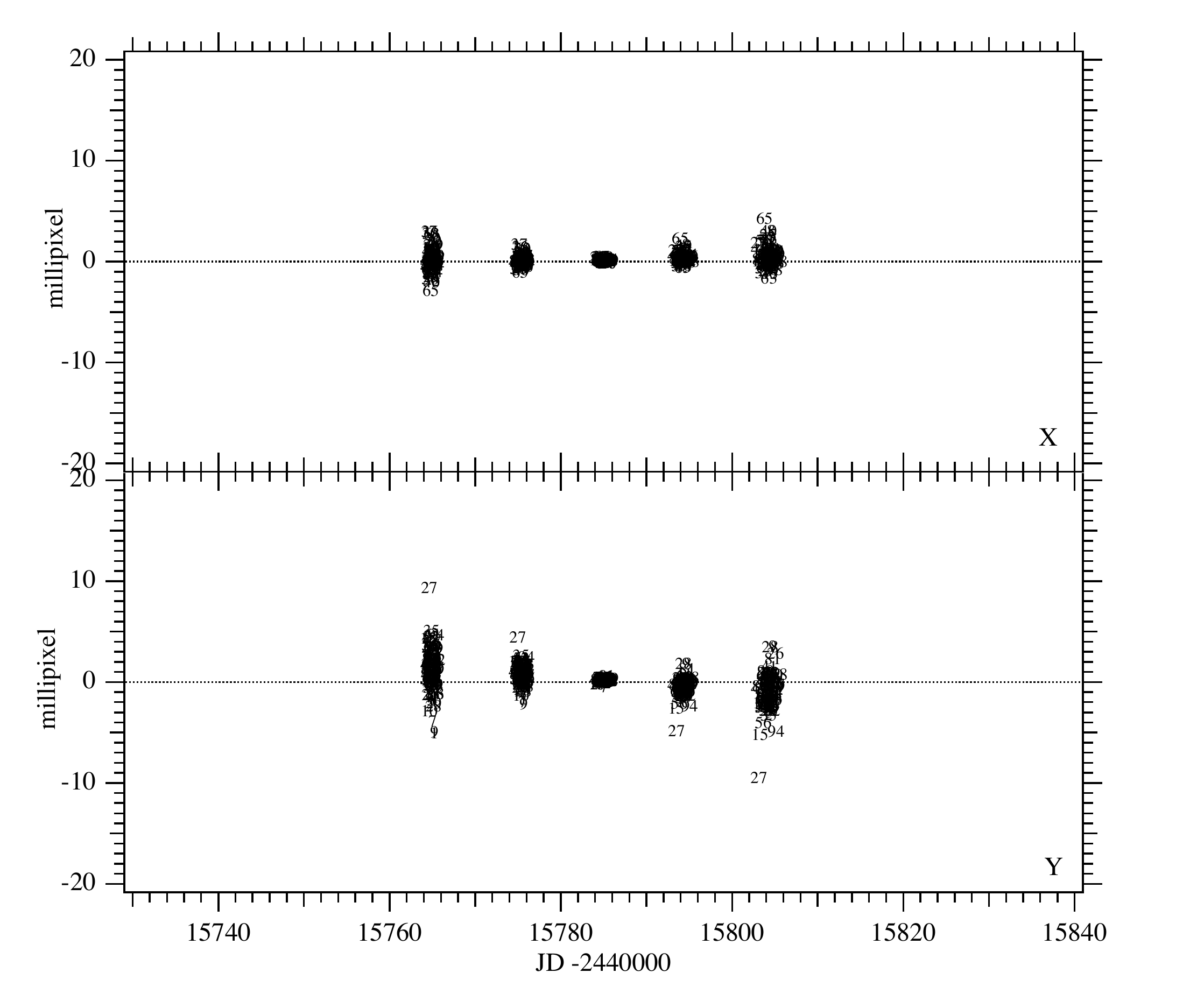}
\caption{x and y residuals as a function of time for the Q10-only four parameter modeling from Section~\ref{WTF1}. The residual clumps from left to right are 'plates' 3--7 first seen in Figure~\ref{Q10r}. Stars are labeled with a running number from 1 to 97. The residuals exhibit far less time-dependency. Stars 9 and 27 continue to exhibit unmodeled behavior.}
\label{Q10ed}
\end{figure}

\begin{figure}
\includegraphics[width=4in]{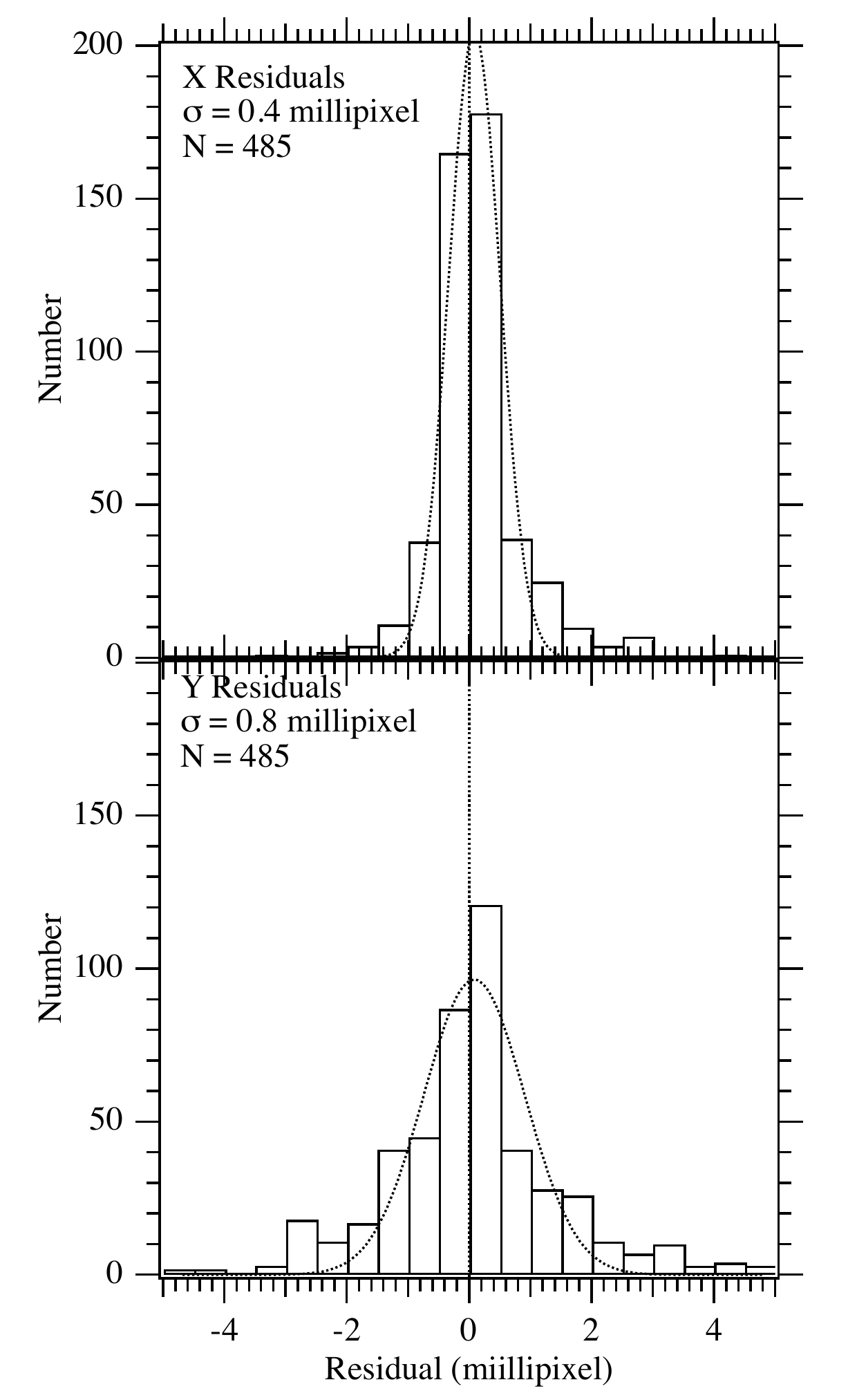}
\caption{Histograms of x and y residuals for the Q10-only four parameter modeling of only plates 3--7 (Figure~\ref{Q10ed}) from Section~\ref{WTF1}. The residuals are well-characterized with Gaussians with 1-$\sigma$ dispersions as indicated.}
\label{Q10H}
\end{figure}

\begin{figure}
\includegraphics[width=6in]{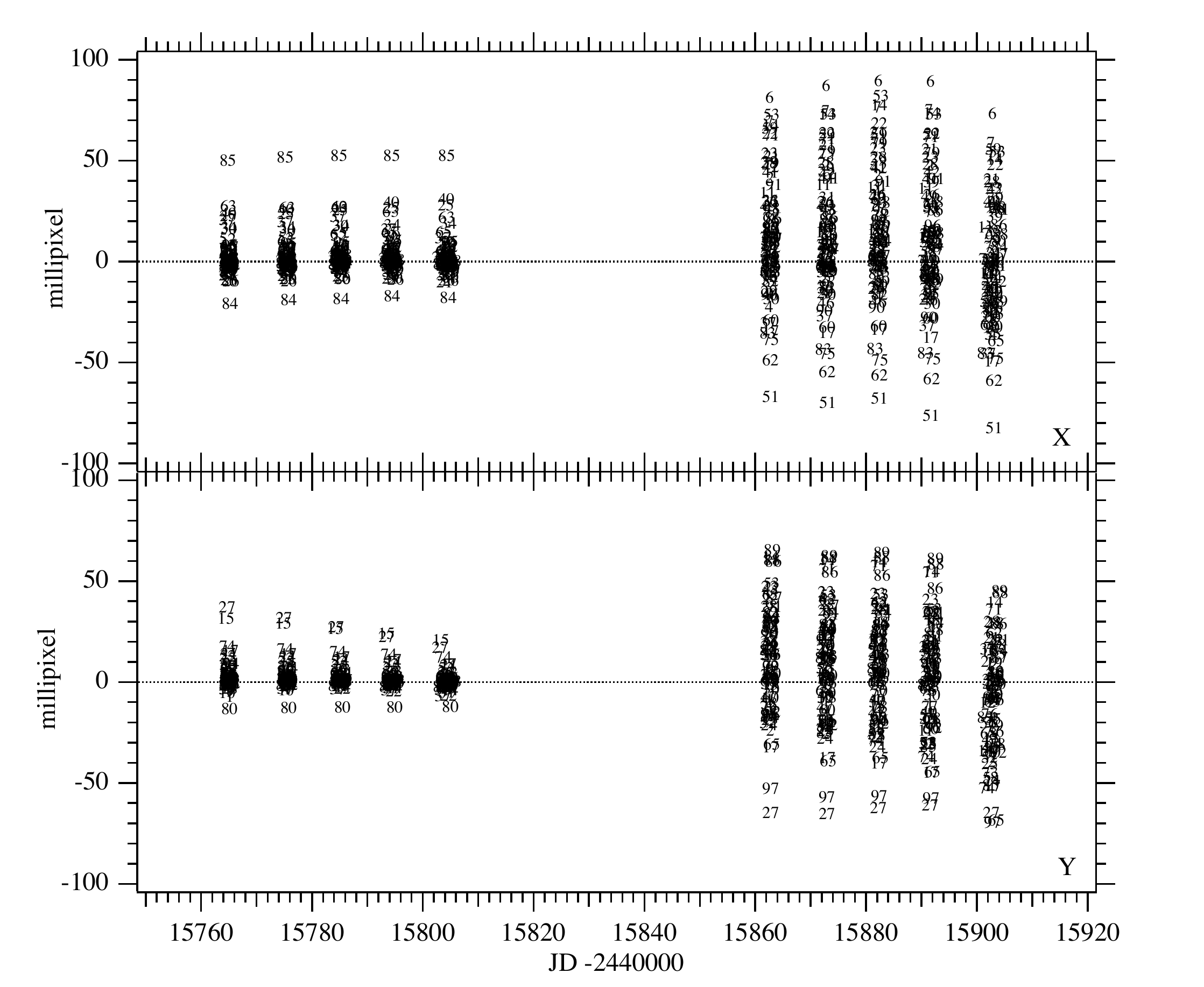}
\caption{x and y residuals in millipixels as a function of time from the full Schmidt 14 parameter modeling from Section~\ref{WTF2}.  The residual clumps on the left hand side (Channel 21) from left to right are 'plates' 3--7. Stars are labeled with a running number from 1 to 97. Note the scale change along the y-axis, a range five times larger than that in Figure~\ref{Q10ed}. The residuals exhibit extreme time-dependency. Star 27 continues to show unmodeled behavior in both channels. Virtually all stars in Channel 37 (right) exhibit unmodeled behavior.}
\label{C21C37}
\end{figure}

\begin{figure}
\includegraphics[width=6in]{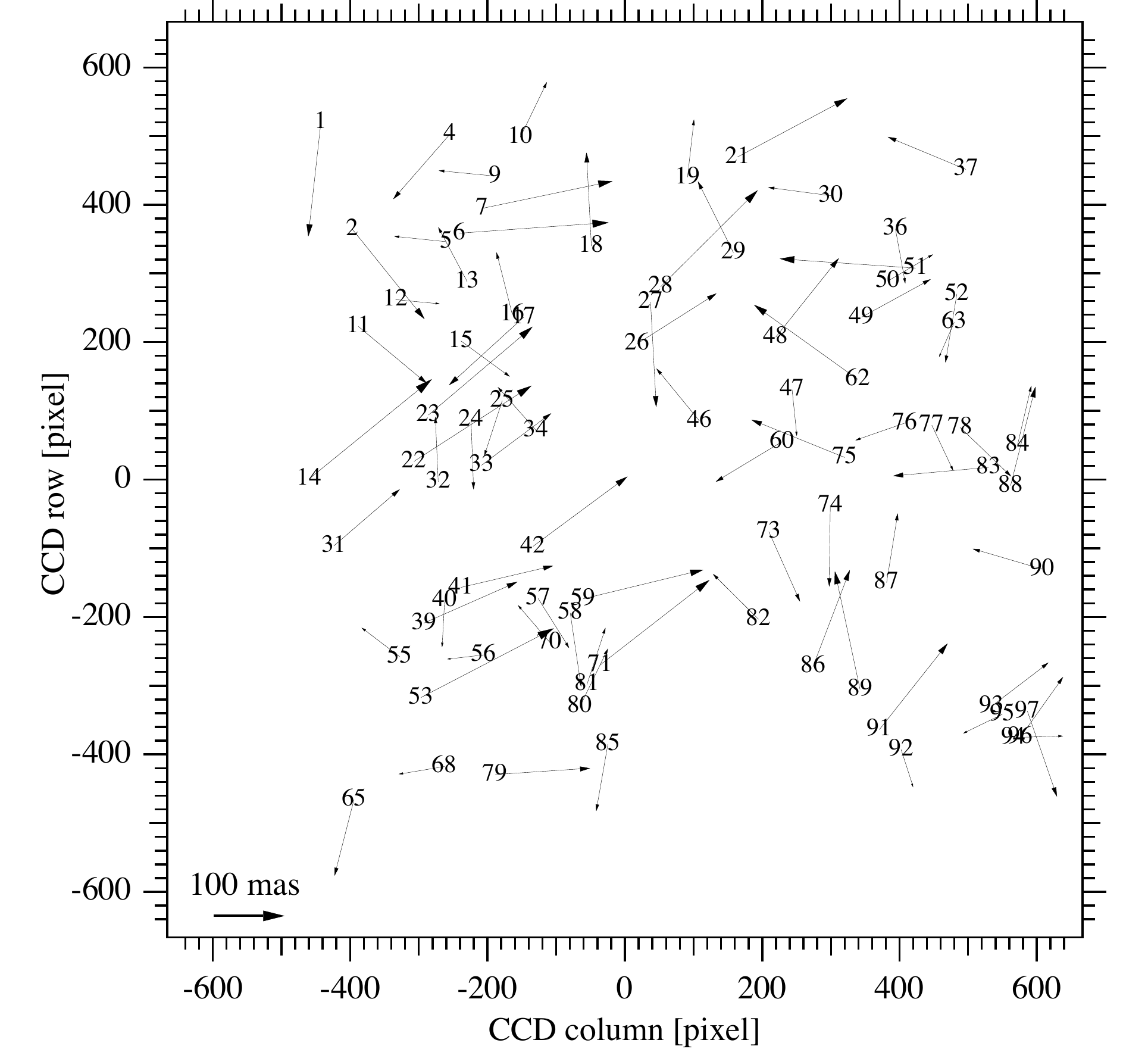}
\caption{Average vector residuals in milliseconds of arc (scale at lower left) as a function of position within Channel 37 from the full Schmidt 14 parameter modeling of Channel 21 and Channel 37 described in Section~\ref{WTF2}. All positions have been re-origined to the CCD center. Note the extreme variation in vector length and position angle over small spatial scales, for example the grouping consisting of stars 22 through 34 (row$\sim$50, column$\sim$-200). Comparing Channel 21 with Channel 37 demonstrates serious astrometric sytematics on very small spatial scales.}
\label{37res}
\end{figure}

\begin{figure}
\includegraphics[width=6in]{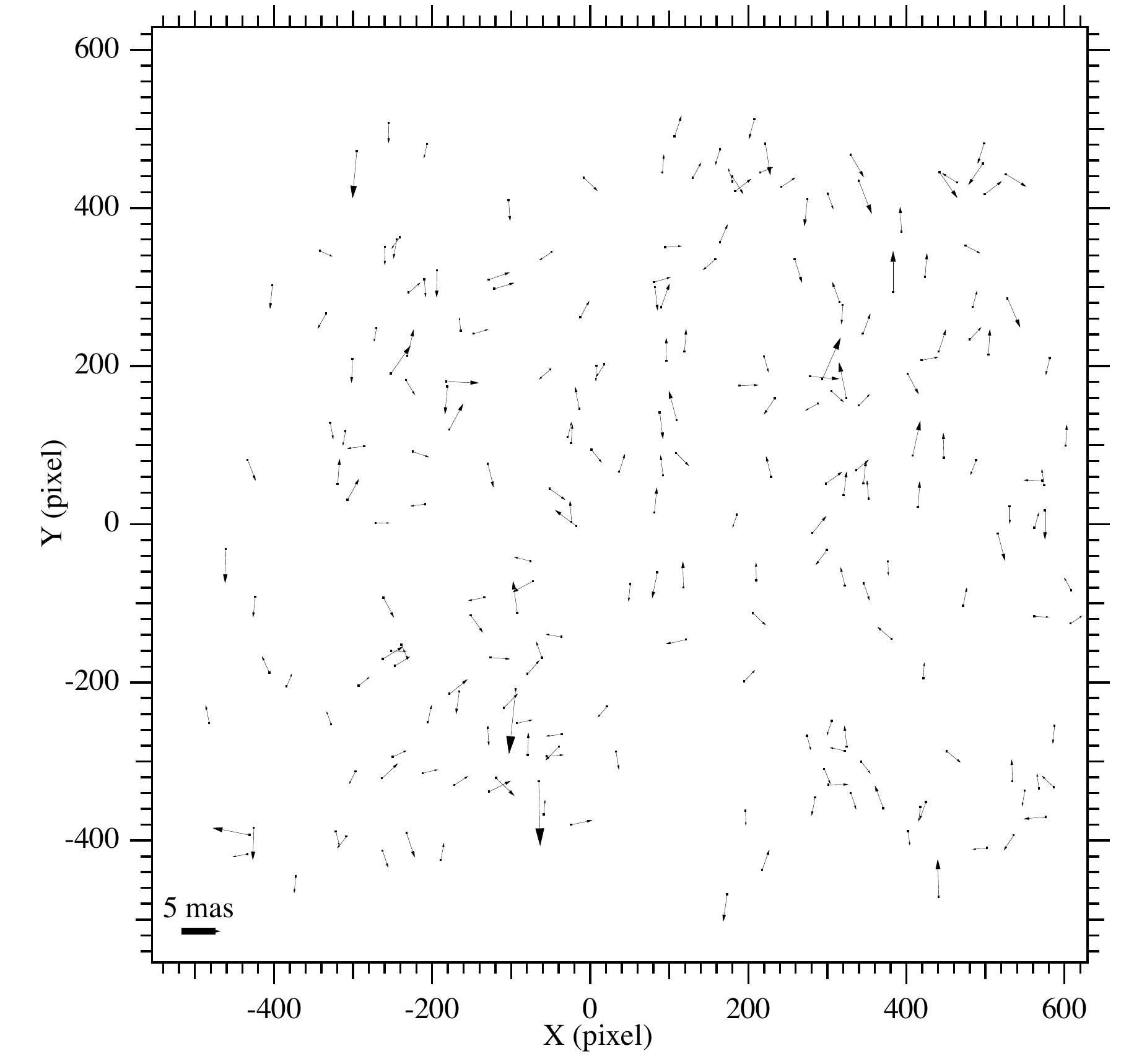}
\caption{Average vector residuals in milliseconds of arc as a function of position within Channel 21 from the full Schmidt 14 parameter modeling of three Season 0 observation sets described in Section~\ref{bmod}.  Other than a strong tendency for larger residuals in the y direction, the pattern is satisfactorily random.}
\label{Ch21res}
\end{figure}

\begin{figure}
\includegraphics[width=6in]{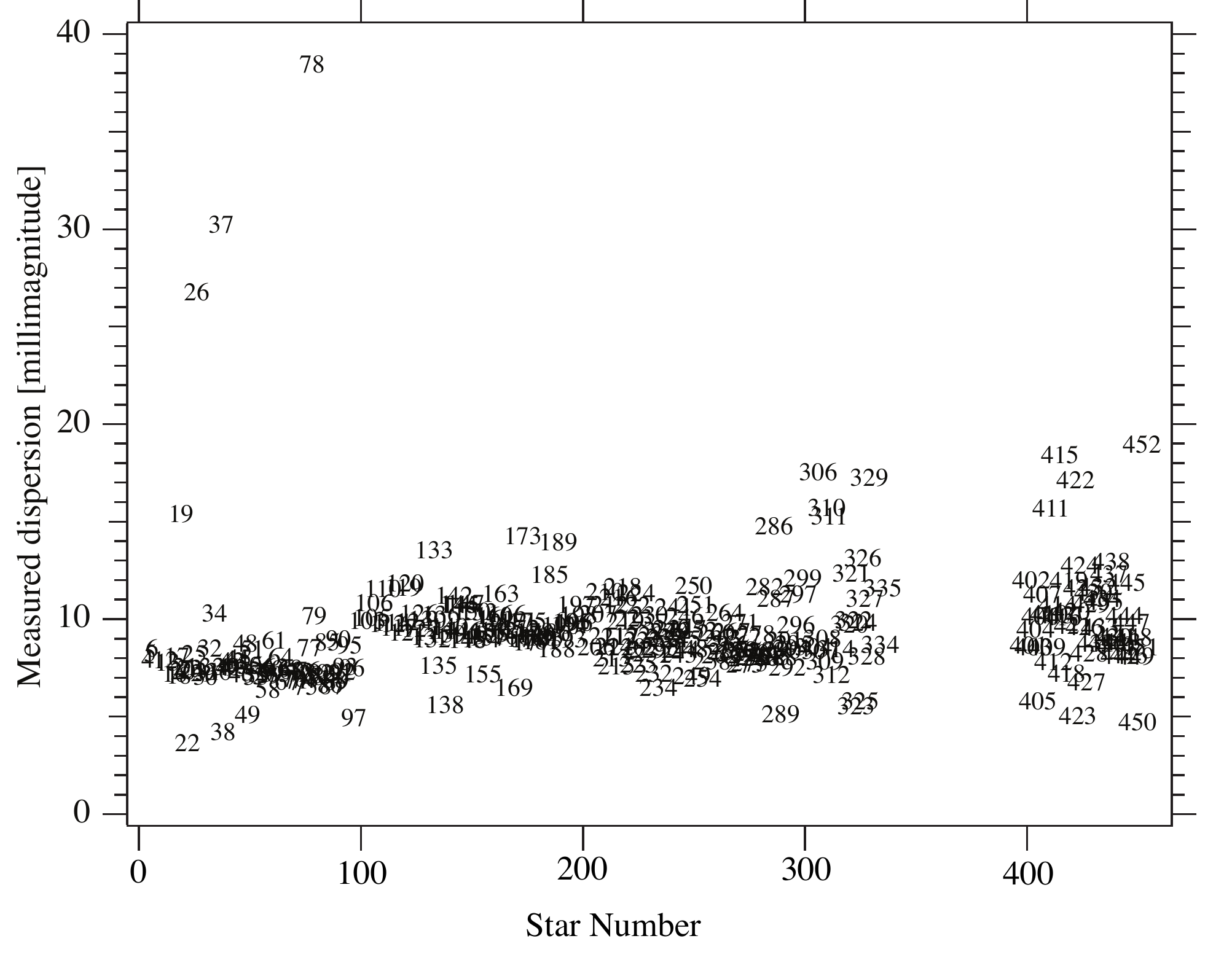}
\caption{Measured photometric dispersion ($m_f$ standard deviation) over 2.1 y with a nine-day cadence for each star modeled in Section~\ref{bmod}. Giants (1--100) exhibit the highest overall variability. Other groups are the hot star sample (101--199), mid-range T$_{\rm eff}$ (201--299), K-M star sample (300--350), and KOI sample (400--452). The trends to smaller photometric variation with number within each sample group (as defined in Section~\ref{DaList}) may be a function of position within Channel 21. Lowest variations are nearer (x,y)=(0,1000), highest nearer (x,y)=(1000,0).}
\label{PhotV}
\end{figure}

\begin{figure}
\includegraphics[width=6in]{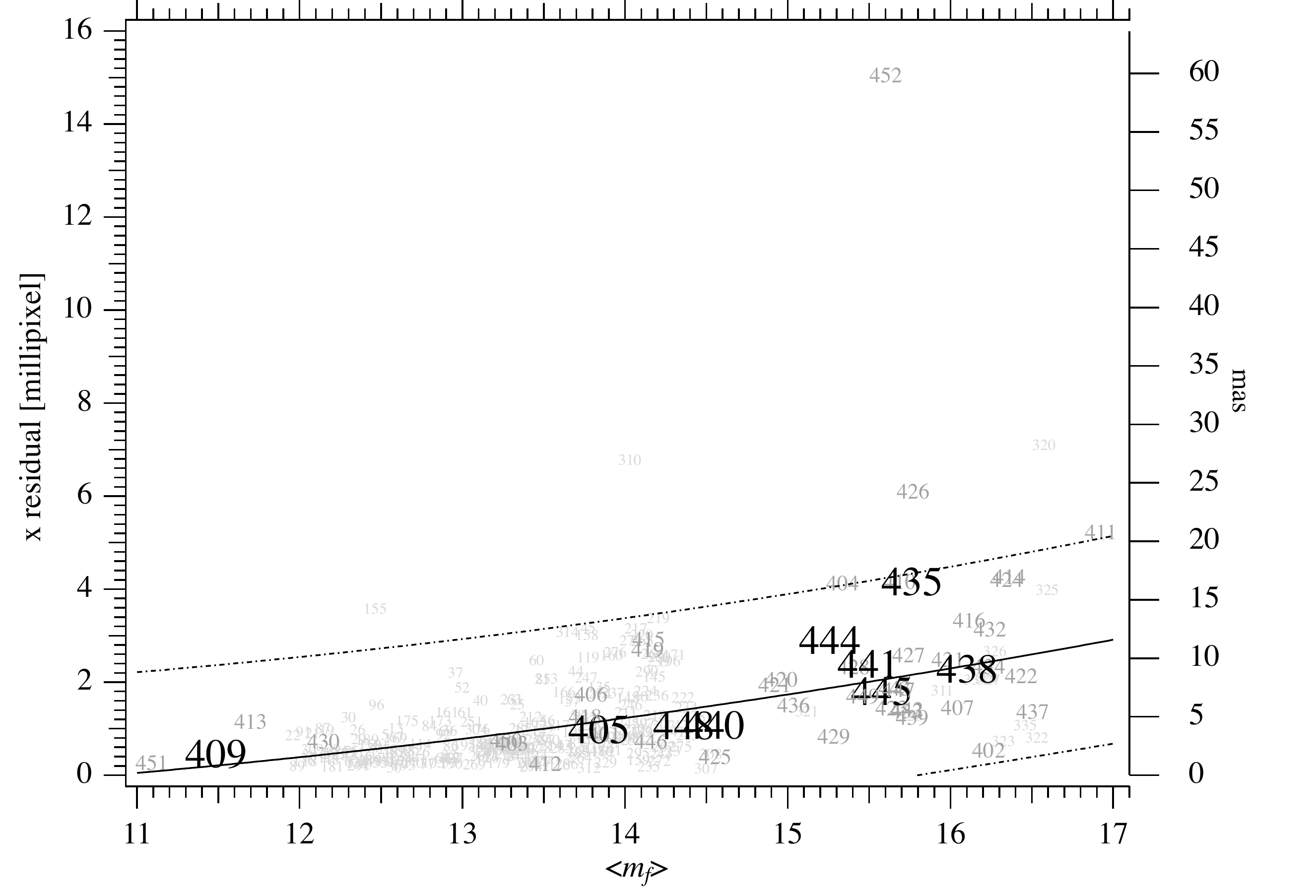}
\caption{Average x residual from the Section~\ref{finpm} modeling plotted against average $m_f$, scale in millipixel on the left, mas on the right. The solid curve is a quadratic fit to average reference star residuals resulting from the 
modeling in Section~\ref{bmod}. Also plotted are upper and lower bounds within which 99\% of the Section~\ref{bmod} reference stars are expected to fall. These stars are plotted with smallest ID numbers. The KOI are plotted with larger symbols. The nine confirmed planetary system host stars are plotted with large bold symbols. All KOI ID numbers are from Table~\ref{KOIM}. Clearly KOI 426 and 452 are astrometrically peculiar, and the nine planetary system host stars behave as expected.}
\label{xA}
\end{figure}

\begin{figure}
\includegraphics[width=6in]{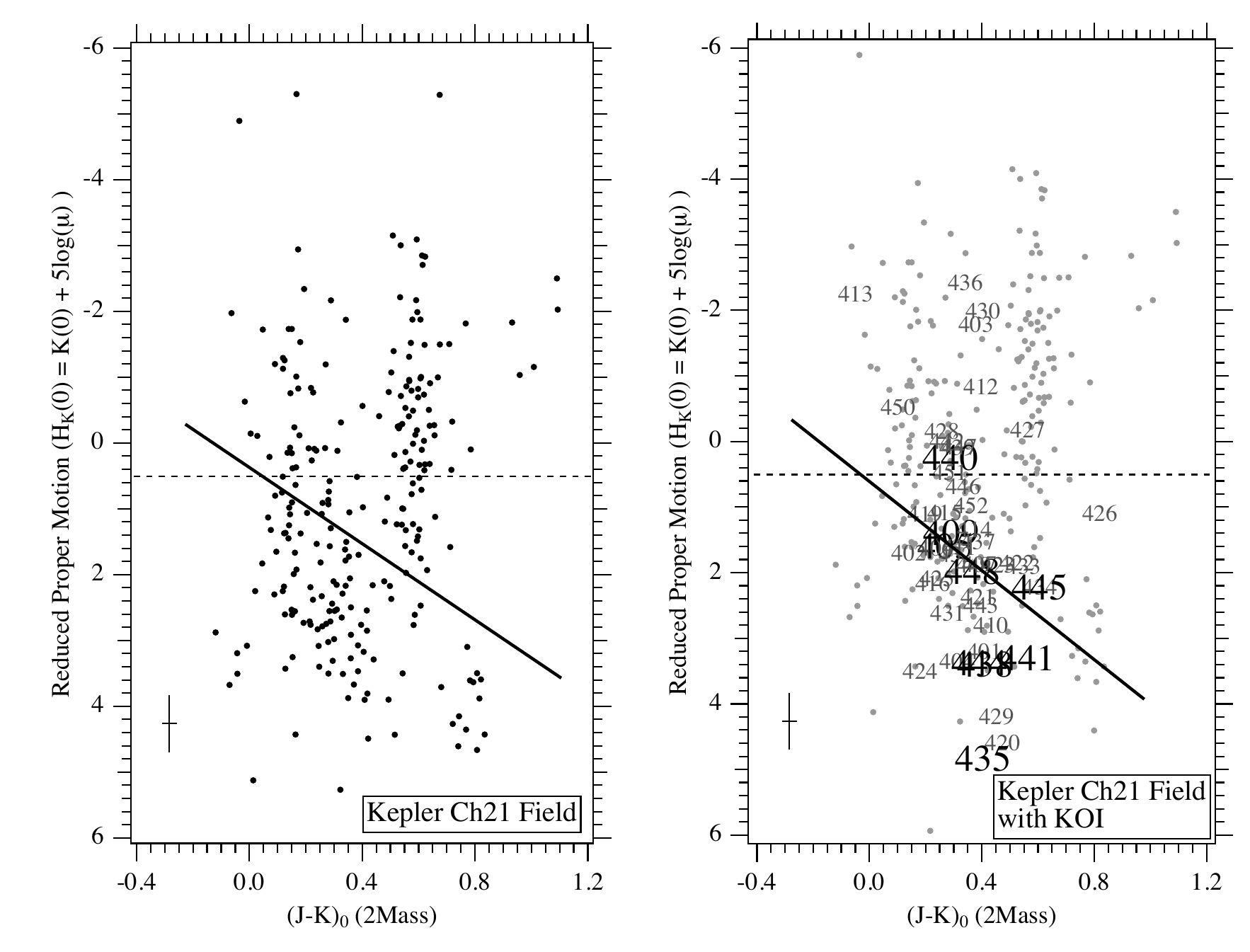}
\caption{Left: RPM from the results of modeling \Kns, UCAC4, and PPMXL data. The average H$_K$(0) and (J-K)$_0$ $\pm1\sigma$ errors are indicated in the lower left. That error is reduced by a factor of three compared to an RPM derived by averaging proper motions from UCAC4 and PPMXL. The heavy tilted line is the location of the main sequence in the Figure~\ref{HR} RPM derived from all sky \HST proper motions. Note the vertical offset in H$_K(0)$ discussed in the text. Right: RPM with a $\Delta$H$_K(0)$ = -1.0 correction, containing the KOI also shifted. Plotted numbers are from Table~\ref{KOIM}. The horizontal dotted line represents a rough demarkation between giant and dwarf stars.}
\label{RPMs}
\end{figure}

 Comparing the \HST - derived RPM (Figure~\ref{HR}, right) with the left-hand RPM in Figure~\ref{RPMs} yields a systematic difference. What we  identify as the locus of main sequence stars from Figure~\ref{HR}, right, appears to be substantially offset towards more negative H$_K$(0) values by $\Delta$H$_K(0)$ = -1.0. As a check we produced an RPM by averaging measured proper motions from the UCAC4 and PPMXL catalogs and obtain the same offset. Averaging the proper motions from the two catalogs, the typical H$_K(0)$ error is 1.58 mag, a factor of three larger than when we include \K astrometry. There is also a far greater increase in error for brighter values of H$_K(0)$.  

To bring the main sequence stars into coincidence with the \HST main sequence would require {decreasing} the average $\mu_{\rm vec}$ proper motions of  the final Table~\ref{KOIpm} proper motions by $\sim10$ \msy. This proper motion offset is likely not from systematic effects on proper motion due to the space velocity of the Sun. The \K field is very near the Solar Apex at RA $\simeq 287\arcdeg$, DEC $\simeq +37\arcdeg$ \citep{Vit13}. With most of the vector of solar motion in the radial direction, stars near the Solar Apex will exhibit very little systematic proper motion due to solar motion. From \cite{Vit13} the average transverse velocity of stars in the solar neighborhood towards the Galactic center is $\bar{U} = 9$ \kms~and the average transverse velocity perpendicular to the Galactic plane $\bar{W}=6$ \kms~for a systematic total velocity V$_t = 10.8$ \kms. Our sample of F--G dwarfs has $\langle K \rangle = 11.85$ mag with  $\langle M_K \rangle \simeq 3$ \citep{Cox00}, hence an average distance of 500pc. The expected proper motion can be estimated from
\beq
\mu_{\rm vec} = \pi V_t/4.74
\eeq
This yields $\mu_{\rm vec}$ = 4.6 \msy, which could explain some but not all the positive H$_K$(0) offset in Figure~\ref{RPMs}.

However as mentioned in Section~\ref{RPM} above, a systematic effect of Galactic rotation on stellar velocities does exist. Our \HST-derived RPM main sequence is composed of stars belonging to the Hyades and Pleiades star clusters. They have a Galactic longitude, $\ell\sim175$\arcdeg. The Kepler field has $\ell\sim74$\arcdeg. The velocity difference due to Galactic rotation is $\sim30$\kms, translating to a proper motion difference, $\Delta\mu_{\rm vec}$ = 12.7 \msy, close to the correction needed above.

We seek luminosity class differentiation, a relative determination within an RPM. We ascribe the need for a correction to bring the Channel 21 RPM into closer agreement with Figure~\ref{HR}, right, to a mix of the random and systematic proper motions just identified. We add the $\Delta$H$_K(0)$ = -1.0 and replot, this time including the KOI, similarly corrected for offset (right-hand side of Figure~\ref{RPMs}).  Identification number in this plot, KIC numbers, and KOI numbers are collected in Table~\ref{KOIM}.

\section{Estimated Absolute Magnitudes for Channel 21 KOI} \label{LC}
\begin{figure}
\includegraphics[width=6in]{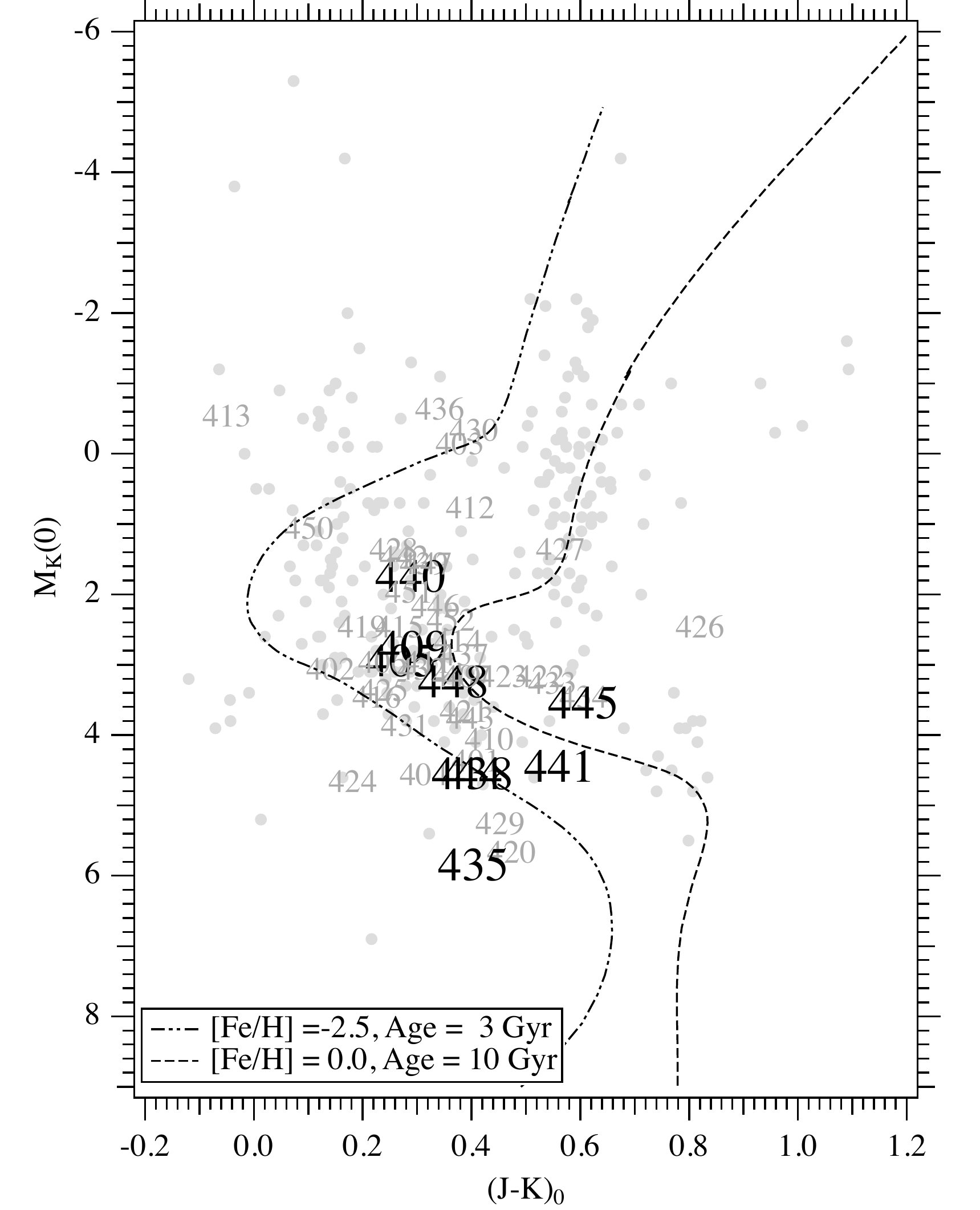}
\caption{An HR diagram for the Section~\ref{finpm} reference stars and the Table~\ref{KOIM} KOIs. M$_K$(0) are calculated using the Figure~\ref{HRRP} linear mapping between M$_K$(0) and H$_K$(0), after applying the Section~\ref{mobet} systematic correction to H$_K$(0). KOI labeling is the same as in Figure~\ref{RPMs}, right.  Lines show predicted loci for 10Gyr age solar metallicity stars (- - -) and 3Gyr age metal-poor stars (-$\cdot\cdot$-) from Dartmouth Stellar Evolution models \citep{Dot08}. Note the large number of KOI sub-giants.}
\label{HRK}
\end{figure}

Table~\ref{KOIM} contains the H$_K(0)$ values (corrected by $\Delta$H$_K(0)$ = -1.0) derived from the Table~\ref{KOIpm} KOI proper motions  and $K(0)$ magnitudes. The listed $K$-band absolute magnitudes, $M_K$(0) are obtained using the Figure~\ref{HRRP} calibration; $M_K$(0) = 1.51$\pm$0.12+(0.90$\pm$0.02)$\times H_K$(0). An $M_K$(0), $(J-K)_0$ HR diagram is shown in Figure~\ref{HRK}.   An HR diagram constructed using H$_K$(0) from the UCAC4, PPMXL average proper motions has a distribution similar to Figure~\ref{HRK}, but with significantly increased scatter. Nine stars in Channel 21 (Season 0) host confirmed planetary systems. Our number 409 is Kepler-100, hosting three confirmed planets \citep{Mar14}, and our number 441 is Kepler-28, hosting two confirmed planets \citep{Ste12}. Recently \cite{Row14} have statistically confirmed a number of multi planet systems. All exoplanet host stars and associated companions found in the Channel 21 field are listed in Table~\ref{XOP}. Most of their positions in the Figure~\ref{HRK} HR diagram lie on or close to a solar-metallicity 10Gyr old main sequence (Dartmouth Stellar Evolution model, Dotter \etal 2008). 
Our number 435 = KOI-1359 has a photometrically-determined (hence, low-precision) lowest metallicity in Table~\ref{XOP}, consistent with sub-dwarf location on an HR diagram, as indicated in Figure~\ref{HRK}. The other eight hosts of confirmed planetary systems have locations in the HR diagram consistent with a main sequence dwarf classification.

Inspection of Figure~\ref{HRK} suggests that a significant number of our astrometric reference stars (and some of the KOI) appear to be sub-giants. As with any apparent magnitude limited survey, the stars observed with \K will have a Malmquist-like bias, i.e., the survey will be biased towards the inclusion of the most luminous objects in the field as a result of the greater volume being surveyed for these intrinsically brighter objects \citep{Mal22}. Therefore, within the \K field there is a significant bias towards observing stars that are more massive and/or more evolved.

The \K target selection attempted to mitigate this bias by selecting stars identified as main-sequence solar-type dwarfs based on their KIC values \citep{Bat10}. However, significant uncertainty in KIC surface gravities make this selection process suspect. \cite{Bro11} concluded that uncertainties in KIC log(g) are $\sim$0.4 dex, and are unreliable for distinguishing giants/main-sequence stars for T$_{\rm eff} \gtrsim$ 5400 K. Consequently, a significant fraction of \K target stars are expected to be F and G spectral type subgiant stars \citep{Far13,Gai13}.

\begin{figure}
\includegraphics[width=6in]{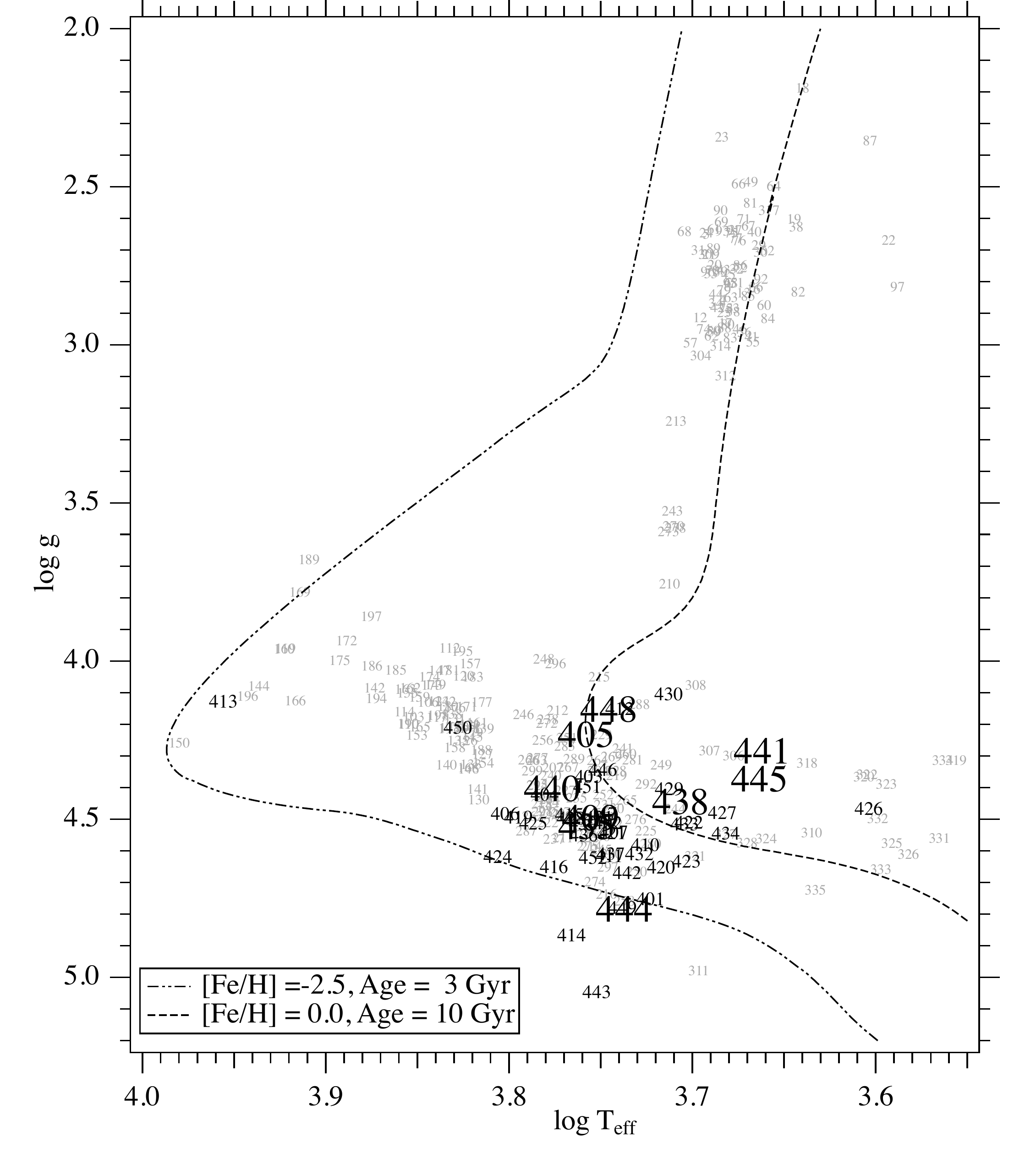}
\caption{A theoretical HR diagram for the Section~\ref{finpm} astrometric reference stars and the Table~\ref{KOIM} KOI. Data are from the MAST. KOI labeling is the same as in Figure~\ref{HRK}. The main sequence and giant branch are apparent. Lines show predicted loci for 10Gyr age solar metallicity stars (- - -) and 3Gyr age metal-poor stars (-$\cdot\cdot$-) from Dartmouth Stellar Evolution models \citep{Dot08}. Compared with Figure~\ref{HRK}, existing log g values apparently do not readily identify sub-giants. }
\label{THR}
\end{figure}

Finally, to argue for the added value of carrying out this astrometry, Figure~\ref{THR} plots a theoretical HR diagram (log g vs. log T) for the astrometric reference stars (grey dots) and the Table~\ref{KOIM} KOI. The log g and T values are from the \K Target Search results tabulated at the STScI MAST\footnote{http://archive.stsci.edu/kepler/kepler\_fov/search.php}. There are very few stars in the expected locus of F--G sub giants.  Cargile \etal (in preparation) have re-measured T$_{\rm eff}$ and log g for 850 KOIs using Keck HiRes archived material and find a sub-giant fraction amongst \K targets similar to that shown in Figure~\ref{HRK}. Again, we include on this plot a range of metallicities and ages from the Dartmouth evolution models.

\section{Summary} \label{SUMM}
\begin{enumerate}
\item Astrometry carried out on  \K data yields significant systematics in position. These systematics correlate with time.

 \item Astrometric performance correlates with \K telescope temperature variations. Larger variations result in poorer astrometry.

\item Astrometric modeling with a previously successful Schmidt model of more than one \K Season fails to produce astrometric precision allowing for the measurement of stellar parallax.

\item Combining \K astrometry for a single Season and Channel and three Quarters with existing catalog positions and proper motions, extends the time baseline to over 12 years. This provides a mapping of the lower-spatial frequency distortions over a Channel, and  improves the precision of measured proper motions to 1.0 \msy,  over a factor of three better than UCAC4 and PPMXL.

\item Applying that astrometric model, \K measurements yield absolute proper motions  for  a number of KOIs with an average proper motion vector error $\sigma_\mu = 2.3 $ \msy, or $\sigma_\mu / \mu= 19.4$\%. In contrast, averaging the UCAC4 and PPMXL  catalog proper motions provide $\sigma_\mu = 5.0 $ \msy, or $\sigma_\mu / \mu= 43.0$\%.

\item An RPM diagram constructed from the proper motions determined by our method, when compared to one based on \HST proper motions, shows a systematic offset. Much of the offset can be attributed to the effect of Galactic rotation on proper motions.

\item The corrected RPM parameter, H$_K$(0), transformed to M$_K$(0) through an H$_K$(0) - M$_K$(0) relation derived from \HST proper motions and parallaxes, yields M$_K$(0) for fifty KOIs, including nine stars with confirmed planetary companions, 8 now confirmed as dwarfs, one a possible sub-dwarf. Six KOIs are identified as giants or sub-giants.
\end{enumerate}

The next significant improvement in KOI proper motions will come from the space-based, all-sky astrometry mission {\it Gaia} \citep{Lin08}  with $\sim20$~microsecond of arc precision proper motions and parallaxes for the brighter KOI's. With parallax there will be no need for RPM diagrams. Final {\it Gaia} results are expected early in the next decade.

\acknowledgments

This paper includes data collected by the \K mission. Funding for  \K  is provided by the NASA Science Mission directorate. All of the \K data presented in this paper were obtained from the Mikulski Archive for Space Telescopes (MAST) at the Space Telescope Science Institute. STScI is operated by the Association of Universities for Research in Astronomy, Inc., under NASA contract NAS5-26555. Support for MAST for non-HST data is provided by the NASA Office of Space Science via grant NNX13AC07G and by other grants and contracts. Direct support for this work was provided to GFB by NASA through grant NNX13AC22G.  Direct support for this work was provided to AMT by NASA through grant NNX12AF76G. PAC acknowledges NSF Astronomy and Astrophysics grant AST-1109612.  This publication makes use of data products from the Two Micron All Sky Survey, which is a joint project of the University of Massachusetts and the Infrared Processing and Analysis Center/California Institute of Technology, funded by NASA and the NSF. This research has made use of the SIMBAD and Vizier databases and Aladin, operated at CDS, Strasbourg, France; the NASA/IPAC Extragalactic Database (NED) which is operated by JPL, California Institute of Technology, under contract with the NASA;  and NASA's Astrophysics Data System Abstract Service.  This research has made use of the NASA Exoplanet Archive, which is operated by the California Institute of Technology, under contract
with the National Aeronautics and Space Administration under the Exoplanet Exploration Program.  Some of the data presented herein were obtained at the W.M. Keck Observatory, which is operated as a scientific partnership among the California Institute of Technology, the University of California and the National Aeronautics and Space Administration. The Observatory was made possible by the generous financial support of the W.M. Keck Foundation. The authors wish to recognize and acknowledge the very significant cultural role and reverence that the summit of Mauna Kea has always had within the indigenous Hawaiian community.  We are most fortunate to have the opportunity to conduct observations from this mountain. GFB thanks Bill Jefferys, Tom Harrison, and Barbara McArthur who over many years contributed to the techniques reported in this paper. GFB and AMT thank Dave Monet for several stimulating discussions that should have warned us off from this project, but didn't. GFB thanks Debra Winegarten for her able assistance, allowing progress on this project. We thank an anonymous referee for a thorough, careful, and useful review which materially improved the final paper.
\clearpage


\bibliography{/Active/myMaster}

\clearpage
\begin{deluxetable}{l l l l l l l l l l}
\tablewidth{7.0in}
\tablecaption{\HST M$_K$(0) and H$_K$(0) \label{THT}} 
\tablehead{\colhead{\#}&
\colhead{ID} &
\colhead{m-M}&
\colhead{M$_K$(0)} &
\colhead{SpT} &
\colhead{$\mu_T$\tablenotemark{a} } &
\colhead{K$_0$} &
\colhead{(J-K)$_0$} &
\colhead{H$_K$(0)} &
\colhead{Ref\tablenotemark{b} } 
}
\startdata
1&HD 213307&7.19&-0.86&B7IV&21.82$\pm$0.42&6.32&-0.12&-1.98$\pm$0.05&B02\\
2&$\upsilon$ AND&0.66&2.20&F8 IV-V&419.26 0.14&2.86&0.32&0.97 0.03&M10\\
3&HD 136118&3.59&2.00&F9V&126.31 1.20&5.60&0.34&1.11 0.03&Mr10\\
4&HD 33636&2.24&3.32&G0V&220.90 0.40&5.56&0.34&2.28 0.03&Ba07\\
5&HD 38529&3.00&1.22&G4IV&162.31 0.11&4.22&0.68&0.27 0.03&B10\\
6&vA 472&3.32&3.69&G5 V&104.69 0.21&7.00&0.50&2.10 0.03&M11\\
7&55 Cnc&0.49&3.49&G8V&539.24 1.18&3.98&0.70&2.64 0.03&SIMBAD\\
8&$\delta$ Cep&7.19&-4.91&F5Iab:&17.40 0.70&2.28&0.52&-6.51 0.09&B07\\
9&vA 645&3.79&4.11&K0V&101.81 0.76&7.90&0.77&2.93 0.03&M11\\
10&HD 128311&1.09&3.99&K1.5V&323.57 0.35&5.08&0.53&2.63 0.03&M13\\
11&$\gamma$ Cep&0.67&0.37&K1IV&189.20 0.50&1.04&0.62&-2.58 0.03&B13\\
12&vA 627&3.31&3.86&K2 V&110.28 0.05&7.17&0.56&2.38 0.03&M11\\
13&$\epsilon$ Eri&-2.47&4.24&K2V&976.54 0.10&1.78&0.45&1.72 0.03&B06\\
14&vA 310&3.43&4.09&K5 V&114.44 0.27&7.52&0.63&2.82 0.03&M11\\
15&vA 548&3.39&4.13&K5 V&105.74 0.01&7.52&0.71&2.64 0.03&M11\\
16&vA 622&3.09&5.13&K7V&107.28 0.05&8.22&0.84&3.38 0.03&M11\\
17&vA 383&3.35&5.01&M1V&102.60 0.32&8.36&0.91&3.42 0.03&M11\\
18&Feige 24&4.17&6.38&M2V/WD&71.10 0.60&10.55&0.69&4.81 0.03&B00a\\
19&GJ 791.2&-0.26&7.57&M4.5V&678.80 0.40&7.31&0.92&6.47 0.03&B00b\\
20&Barnard&-3.68&8.21&M4Ve&10370.00 0.30&4.52&0.72&9.60 0.03&B99\\
21&Proxima&-4.43&8.81&M5Ve&3851.70 0.10&4.38&0.97&7.31 0.03&B99\\
22&TV Col&7.84&4.84&WD&27.72 0.22&12.68&0.49&4.89 0.03&M01\\
23&DeHt5&7.69&7.84&WD&21.93 0.12&15.53&-0.07&7.24 0.03&B09\\
24&N7293&6.67&7.87&WD&38.99 0.24&14.54&-0.23&7.49 0.03&B09\\
25&N6853&8.04&2.54&WD&8.70 0.11&10.58&1.13&0.27 0.04&B09\\
26&A31&8.97&6.69&WD&10.49 0.13&15.66&0.25&5.77 0.04&B09\\
27&V603 Aql&7.20&4.12&CNe&15.71 0.19&11.32&0.31&2.30 0.04&H13\\
28&DQ Her&8.06&5.00&CNe&13.47 0.32&13.06&0.46&3.70 0.06&H13\\
29&RR Pic&8.71&3.54&CNe&5.21 0.36&12.25&0.18&0.83 0.15&H13\\
30&HP Lib&6.47&7.35&WD&33.59 1.54&13.82&-0.12&6.45 0.10&R07\\
31&CR Boo&7.64&8.59&WD&38.80 1.78&16.23&-1.52&9.17 0.10&R07\\
32&V803 Cen&7.70&6.12&WD&9.94 2.98&13.82&-0.10&3.81 0.65&R07\\
33&$\ell$ Car&8.56&-7.55&G3Ib&15.20 0.50&0.99&0.55&-8.10 0.08&B07\\
34&$\zeta$ Gem&7.81&-5.73&G0Ibv&6.20 0.50&2.13&0.23&-8.91 0.18&B07\\
35&$\beta$ Dor&7.50&-5.62&F6Ia&12.70 0.80&2.06&0.48&-7.42 0.14&B07\\
36&FF Aql&7.79&-4.39&F6Ib&7.90 0.80&3.45&0.40&-7.06 0.22&B07\\
37&RT Aur&8.15&-4.25&F8Ibv&15.00 0.40&3.90&0.28&-5.22 0.06&B07\\
38&$\kappa$ Pav&6.29&-3.52&F5Ib-II:&18.10 0.10&2.71&0.62&-6.00 0.03&B11\\
39&VY Pyx&6.00&-0.26&F4III&31.80 0.20&5.63&0.33&-1.86 0.03&B11\\
40&P3179&5.65&3.02&G0V:&50.36 0.40&8.67&0.35&2.18 0.03&S05\\
41&P3063&5.65&4.68&K6V:&45.30 0.50&10.33&0.82&3.61 0.04&S05\\
42&P3030&5.65&4.97&K9V:&43.20 0.50&10.62&0.83&3.79 0.04&S05\\
\enddata
\tablenotetext{a}{$\mu_T = (\mu_{\rm RA}^2 + \mu_{\rm DEC}^2)^{1/2}$ in \msy}
\tablenotetext{b}{B99=\cite{Ben99}, B00a=\cite{Ben00a}, B00b=\cite{Ben00b}, B02=\cite{Ben02b}, B06=\cite{Ben06}, B07=\cite{Ben07}, B09=\cite{Ben09}, B11=\cite{Ben11}, Ba07=\cite{Bea07}, H13=\cite{Har13}, Mr10=\cite{Mar10}, M01=\cite{McA01}, M11=\cite{McA11},  R07=\cite{Roe07}, S05=\cite{Sod05}}
\end{deluxetable}

\begin{deluxetable}{l l l l l}
\tablewidth{5.0in}
\tablecaption{Companion Test Stars \label{Textern}} 
\tablehead{\colhead{\#\tablenotemark{a}}&
\colhead{KID} &
\colhead{KEPMAG}&
\colhead{K} &
\colhead{FluxFrac\tablenotemark{b}} 
}
\startdata
4&5698236&15.637&14.243&0.886\\
5&5698325&12.264&10.747&0.920\\
10&5698466&13.113&11.561&0.921\\
34&5783576&14.135&12.656&0.874\\
62&5784222&15.475&13.885&0.881\\
67&5784291&13.148&11.074&0.936\\
77&5869153&15.596&13.537&0.706\\
96&5869586&15.466&13.672&0.886\\
103&5869826&15.768&13.473&0.774\\
106&5870047&11.747&6.328&0.962
\enddata
\tablenotetext{a}{Numbering in Figure~\ref{GBres}}
\tablenotetext{b}{Fraction of target flux in the \K project-defined optimum aperture.}
\end{deluxetable}

 
\begin{deluxetable}{l l l l r r r}
\tablewidth{7.0in}
\tablecaption{Channel 21 KOI Proper Motions $(\mu)$ \label{KOIpm}} 
\tablehead{\colhead{ID}&
\colhead{KID} &
\colhead{KOI}&
\colhead{$m_f$} &
\colhead{$\mu_{\rm RA}$\tablenotemark{a}} &
\colhead{$\mu_{\rm DEC}$\tablenotemark{b}} &
\colhead{$\mu_{\rm T} $} 
}
\startdata
401&6362874&1128&13.51&-0.0063$\pm$0.0008&-0.0307$\pm$0.0008&0.0314$\pm$0.0011\\
402&6364215&2404&15.66&0.0005 0.0015&-0.0056 0.0019&0.0056 0.0024\\
403&6364582&3456&12.99&0.0022 0.0011&0.0029 0.0005&0.0036 0.0012\\
404&6441738&1246&14.90&-0.0038 0.0013&0.0158 0.0012&0.0163 0.0017\\
405&6442340&664&13.48&0.0090 0.0009&-0.0103 0.0011&0.0137 0.0015\\
406&6442377&176&13.43&0.0079 0.0009&0.0096 0.0012&0.0124 0.0015\\
407&6520519&4749&15.61&0.0026 0.0017&-0.0061 0.0019&0.0067 0.0025\\
408&6520753&4504&11.20&0.0046 0.0087&-0.0535 0.0061&0.0537 0.0106\\
409&6521045&41&15.20&0.0205 0.0003&-0.0275 0.0006&0.0343 0.0007\\
410&6522242&855&15.98&0.0025 0.0037&0.0128 0.0029&0.0130 0.0047\\
411&6523058&4549&13.16&0.0041 0.0019&0.0024 0.0017&0.0048 0.0025\\
412&6523351&3117&11.38&0.0052 0.0006&-0.0021 0.0006&0.0056 0.0009\\
413&6603043&368&15.90&-0.0029 0.0005&-0.0018 0.0004&0.0034 0.0006\\
414&6604328&1736&13.80&0.0033 0.0026&0.0026 0.0021&0.0042 0.0034\\
415&6605493&2559&15.55&-0.0052 0.0011&-0.0076 0.0012&0.0092 0.0016\\
416&6606438&2860&13.42&0.0044 0.0025&0.0051 0.0020&0.0068 0.0032\\
419&6607447&1242&13.75&0.0087 0.0009&0.0002 0.0014&0.0087 0.0017\\
420&6607644&4159&14.50&-0.0070 0.0017&0.0395 0.0018&0.0402 0.0025\\
421&6690082&1240&14.47&-0.0027 0.0010&-0.0130 0.0012&0.0133 0.0015\\
422&6690171&3320&15.95&0.0060 0.0021&-0.0027 0.0015&0.0066 0.0026\\
423&6690836&2699&15.23&-0.0081 0.0031&-0.0026 0.0019&0.0085 0.0036\\
424&6691169&4890&15.77&-0.0006 0.0016&-0.0105 0.0021&0.0106 0.0026\\
425&6693640&1245&14.20&0.0098 0.0012&0.0059 0.0013&0.0115 0.0018\\
426&6773862&1868&15.22&-0.0089 0.0024&-0.0014 0.0020&0.0091 0.0031\\
427&6774537&2146&15.33&-0.0023 0.0014&0.0026 0.0013&0.0035 0.0019\\
428&6774880&2062&15.00&0.0024 0.0019&-0.0019 0.0016&0.0031 0.0025\\
429&6776401&1847&14.81&-0.0031 0.0014&-0.0295 0.0015&0.0297 0.0020\\
430&6779260&2678&11.80&0.0000 0.0006&0.0063 0.0004&0.0063 0.0007\\
431&6779726&3375&15.70&0.0011 0.0031&-0.0084 0.0024&0.0085 0.0040\\
432&6862721&1982&15.77&0.0026 0.0017&0.0051 0.0022&0.0057 0.0027\\
433&6863998&867&15.22&0.0076 0.0014&0.0054 0.0012&0.0093 0.0019\\
434&6945786&3136&15.74&0.0088 0.0018&0.0010 0.0019&0.0089 0.0026\\
435&6946199&1359&15.23&0.0291 0.0032&-0.0082 0.0023&0.0303 0.0040\\
436&6947164&3531&14.62&-0.0002 0.0009&0.0010 0.0013&0.0011 0.0016\\
437&6947668&3455&15.80&-0.0027 0.0015&-0.0046 0.0017&0.0054 0.0023\\
438&6948054&869&15.60&0.0112 0.0023&0.0097 0.0020&0.0148 0.0031\\
439&6948480&2975&15.31&-0.0030 0.0013&-0.0004 0.0013&0.0030 0.0018\\
440&6949061&1960&14.13&0.0050 0.0012&-0.0022 0.0010&0.0055 0.0015\\
441&6949607&870&15.04&-0.0003 0.0017&0.0216 0.0019&0.0216 0.0026\\
442&6949898&3031&15.27&-0.0020 0.0021&-0.0021 0.0015&0.0029 0.0026\\
443&7031517&871&15.22&0.0066 0.0021&-0.0072 0.0014&0.0097 0.0025\\
444&7032421&1747&14.79&0.0088 0.0020&0.0160 0.0026&0.0182 0.0033\\
445&7033233&2339&15.13&0.0102 0.0016&0.0072 0.0017&0.0125 0.0023\\
446&7033671&670&13.77&0.0034 0.0008&-0.0076 0.0007&0.0083 0.0011\\
447&7115291&3357&15.19&0.0030 0.0018&-0.0014 0.0013&0.0034 0.0023\\
448&7115785&672&14.00&-0.0078 0.0011&-0.0117 0.0012&0.0141 0.0016\\
449&7118364&873&15.02&0.0072 0.0018&-0.0040 0.0015&0.0083 0.0023\\
450&7199060&4152&12.97&-0.0047 0.0010&-0.0028 0.0005&0.0054 0.0011\\
451&7199397&75&10.78&-0.0019 0.0007&0.0265 0.0006&0.0266 0.0009\\
452&7199906&1739&15.13&0.0029 0.0014&0.0041 0.0018&0.0050 0.0022\\
\enddata
\tablenotetext{a}{Proper motions in seconds of arc per year.}
\tablenotetext{b}{Corrected for cos$\delta$ declination.}
\end{deluxetable}


\begin{deluxetable}{l l r l l c c l}
\tablewidth{7.0in}
\tablecaption{Channel 21 KOI Absolute Magnitude \label{KOIM}} 
\tablehead{\colhead{ID}&
\colhead{KID} &
\colhead{KOI}&
\colhead{K$_0$} &
\colhead{(J-K)$_0$} &
\colhead{H$_K$(0)\tablenotemark{a}} &
\colhead{M$_K$\tablenotemark{b} } &
\colhead{Status\tablenotemark{c} } 
}
\startdata
401&6362874&1128&11.75&0.41&3.23$\pm$0.08&4.4$\pm$0.1&PC\\
402&6364215&2404&14.02&0.14&1.73 0.96&3.1 0.4&PC\\
403&6364582&3456&11.38&0.38&-1.76 0.71&-0.1 0.9&PC\\
404&6441738&1246&13.34&0.31&3.39 0.23&4.6 0.1&PC\\
405&6442340&664&11.96&0.27&1.65 0.23&3.0 0.1&Kepler-206b,c,d\\
406&6442377&176&12.18&0.24&1.65 0.26&3.0 0.1&PC\\
407&6520519&4749&13.77&0.39&1.90 0.81&3.2 0.3&PC\\
408&6520753&4504&13.94&0.43&6.59 0.43&7.4 0.1&PC\\
409&6521045&41&9.76&0.29&1.43 0.05&2.8 0.1&Kepler-100b,c,d\\
410&6522242&855&13.27&0.43&2.83 0.78&4.1 0.2&PC\\
411&6523058&4549&14.24&0.32&1.64 1.14&3.0 0.4&PC\\
412&6523351&3117&11.47&0.40&-0.80 0.34&0.8 1.7&PC\\
413&6603043&368&11.03&-0.05&-2.22 0.39&-0.5 0.3&PC\\
414&6604328&1736&14.24&0.38&1.37 1.73&2.7 0.7&PC\\
415&6605493&2559&12.30&0.27&1.12 0.38&2.5 0.2&PC\\
416&6606438&2860&14.04&0.23&2.19 1.03&3.5 0.3&PC\\
419&6607447&1242&12.43&0.20&1.14 0.42&2.5 0.2&PC\\
420&6607644&4159&12.60&0.47&4.62 0.14&5.7 0.1&PC\\
421&6690082&1240&12.81&0.39&2.41 0.26&3.7 0.1&PC\\
422&6690171&3320&13.79&0.53&1.88 0.85&3.2 0.3&EB; PC\\
423&6690836&2699&13.27&0.46&1.91 0.92&3.2 0.3&PC\\
424&6691169&4890&14.41&0.18&3.53 0.54&4.7 0.1&PC\\
425&6693640&1245&12.79&0.24&2.12 0.33&3.4 0.1&PC\\
426&6773862&1868&12.27&0.82&1.12 0.70&2.5 0.4&PC\\
427&6774537&2146&13.13&0.56&-0.15 1.16&1.4 1.3&PC\\
428&6774880&2062&13.45&0.26&-0.12 1.75&1.4 2.0&PC\\
429&6776401&1847&12.88&0.45&4.23 0.15&5.3 0.1&PC\\
430&6779260&2678&10.07&0.41&-1.96 0.23&-0.3 0.3&PC\\
431&6779726&3375&14.00&0.28&2.65 1.01&3.9 0.3&PC\\
432&6862721&1982&13.97&0.31&1.76 1.03&3.1 0.4&PC\\
433&6863998&867&13.11&0.55&1.94 0.44&3.3 0.2&PC\\
434&6945786&3136&13.52&0.61&2.24 0.63&3.5 0.2&PC\\
435&6946199&1359&13.48&0.40&4.88 0.28&5.9 0.1&R14\\
436&6947164&3531&13.06&0.34&-2.39 2.66&-0.6 1.8&EB; PC\\
437&6947668&3455&13.93&0.39&1.56 0.92&2.9 0.4&PC\\
438&6948054&869&13.59&0.41&3.45 0.45&4.6 0.1&Kepler-245b,c,d\\
439&6948480&2975&13.69&0.31&0.12 1.28&1.6 1.2&PC\\
440&6949061&1960&12.64&0.29&0.30 0.62&1.8 0.5&Kepler-343b,c\\
441&6949607&870&12.68&0.56&3.35 0.26&4.5 0.1&Kepler-28b,c\\
442&6949898&3031&13.61&0.28&0.02 1.83&1.5 2.1&PC\\
443&7031517&871&13.60&0.40&2.53 0.56&3.8 0.2&PC\\
444&7032421&1747&13.10&0.39&3.43 0.39&4.6 0.1&R14\\
445&7033233&2339&12.78&0.61&2.27 0.40&3.6 0.1&R14\\
446&7033671&670&12.15&0.33&0.72 0.28&2.2 0.2&PC\\
447&7115291&3357&13.49&0.32&0.11 1.45&1.6 1.3&EB; PC\\
448&7115785&672&12.32&0.37&2.04 0.25&3.3 0.1&Kepler-209b,c\\
449&7118364&873&13.34&0.37&1.91 0.62&3.2 0.2&PC\\
450&7199060&4152&11.84&0.10&-0.49 0.43&1.1 0.8&PC\\
451&7199397&75&9.37&0.29&0.50 0.08&2.0 0.1&PC\\
452&7199906&1739&13.43&0.36&1.01 0.93&2.4 0.5&PC\\
\enddata
\tablenotetext{a}{H$_K$(0) = K(0) + 5log($\mu_{\rm T}$) with $\Delta$H$_K(0)$=-1.0 correction.}
\tablenotetext{b}{M$_K$(0) = 1.51$\pm$0.12 + (0.90$\pm$0.02)$\times$H$_K$(0) from the Figure~\ref{HRRP} calibration.}
\tablenotetext{c}{PC = Planetary candidate, EB = Eclipsing Binary, FP = False Positive, Kepler = designated exoplanets have been confirmed, R14 = statistical multi-exoplanet confirmation \citep{Row14}.}

\end{deluxetable}

\begin{landscape}
\begin{deluxetable}{l l l l l c c l l l l l l}
\tablewidth{7.5in}
\tablecaption{Channel 21, Season 0 Planetary Systems \label{XOP}} 
\tablehead{\colhead{ID}&
\colhead{KID} &
\colhead{KOI}&
\colhead{Planet\tablenotemark{a}} &
\colhead{P(days)}&
\colhead{ref.\tablenotemark{b}}&
\colhead{T$_{\rm eff}$}&
\colhead{R$_\odot$}&
\colhead{log g}&
\colhead{[Fe/H]}&
\colhead{(J-K)$_0$} &
\colhead{M$_K$(0) } }
\startdata
405&6442340&664.01&206c&13.1375&R14&5764&1.19&4.24&-0.15&0.27&3.0$\pm$0.1\\
&&664.02&206b&7.7820&&&&&&&&\\
&&664.03&206d&23.4428&&&&&&&&\\
409&6521045&41.01&100b&12.816&{Mar14}&5825&1.49&4.13&+0.02&0.29&2.8 0.1\\
&&41.02&100c&6.887&&&&&&&&\\
&&41.03&100d&35.333&&&&&&&&\\
435&6946199&1359.01&&37.101&R14&5985&0.85&4.53&-0.51&0.4&5.9 0.1&\\
&&1359.02&&104.8202&&&&&&&&\\
438&6948054&869.01&245b&7.4902&R14&5100&0.80&4.56&-0.03&0.41&4.6 0.1&\\
&&869.02&245d&36.2771&&&&&&&&\\
&&869.03&245c&17.4608&&&&&&&&\\
440&6949061&1960.01&343b&8.9686&R14&5807&1.43&4.18&-0.14&0.29&1.8 0.5\\
&&1960.02&343c&23.2218&&&&&&&&\\
441&6949607&870.01&28b&5.9123&R14&4633&0.67&4.65&+0.34&0.56&4.5 0.1\\
&&870.02&28c&8.9858&Ste12&&&&&&&\\
444&7032421&1747.01&&20.5585&R14&5658&0.89&4.54&+0.07&0.39&4.6 0.1\\
&&1747.02&&0.5673&&&&&&&&\\
445&7033233&2339.01&&2.0323&R14&4666&0.68&4.64&+0.38&0.61&3.5 0.1&\\
&&2339.02&&65.1900&&&&&&&&\\
448&7115785&672.01&209b&41.7499&R14&5513&0.94&4.47&+0.01&0.37&3.4 0.1\\
\enddata
\tablenotetext{a}{Assigned number in the Kepler confirmed planet sequence, e.g., Kepler-206c.}
\tablenotetext{a}{Mar14=\cite{Mar14}, Ste12=\cite{Ste12}, R14=\cite{Row14}.}
\end{deluxetable}
\end{landscape}
%
%

\clearpage

\clearpage

\clearpage
\clearpage

\clearpage
\clearpage

\clearpage
\clearpage

\clearpage

\clearpage

\clearpage

\clearpage

\clearpage

\clearpage

\clearpage

\clearpage

\clearpage

\clearpage

\clearpage

\clearpage

\clearpage

\end{document}